\documentclass[fleqn,usenatbib]{mnras}
\usepackage[T1]{fontenc}
\usepackage{ae,aecompl}

\usepackage[UKenglish]{babel}
\usepackage[babel=true]{csquotes}  % Correct quote -- \enquote

\usepackage[final]{microtype}
\usepackage{newtxtext} % font
\usepackage{amsmath}	% Advanced maths commands -- align
\usepackage{newtxmath} % font
\usepackage{amssymb}	% Extra maths symbols -- \gtrsim
\usepackage{upgreek}	% \uppi
\usepackage{gensymb}  % \degree

\usepackage{graphicx}	% Including figure files

\usepackage{subfig}
\usepackage{booktabs}	% Table formatting
\usepackage{multirow}	% Have value span multiple rows -- \multirow{num rows}{width}{text}
\usepackage[capitalise, nameinlink, noabbrev]{cleveref}
\creflabelformat{equation}{#2#1#3}	% To remove brackets around equation numbers

\hypersetup{pdfencoding=unicode,}

\usepackage{etoolbox}
\makeatletter
\patchcmd\@combinedblfloats{\box\@outputbox}{\unvbox\@outputbox}{}{%
}%
\makeatother
\newcommand{\img}[1]{./img/#1.pdf}

\newcommand{\wasyfamily}{\fontencoding{U}\fontfamily{wasy}\selectfont}
\newcommand{\NoExt}{\hbox{\tiny \wasyfamily\symbol{35}}}
\newcommand{\Ext}{\hbox{\tiny \wasyfamily\symbol{32}}}
\newcommand{\ExtNoExt}{\hbox{\tiny \wasyfamily\symbol{71}\llap{\wasyfamily\symbol{35}}}}
\newcommand{\relDur}{\ensuremath{R_{\ExtNoExt{}}}}

\title[SFR tracer lifetimes with extinction]{\vspace{-4mm}An uncertainty principle for star formation -- V.\ The influence of dust extinction on star formation rate tracer lifetimes and the inferred molecular cloud lifecycle\vspace{-4mm}}

\author[Daniel~T.~Haydon et al.]{%
	\vspace{-1mm}Daniel~T.~Haydon,\textsuperscript{1}
	Yusuke~Fujimoto,\textsuperscript{2,3}
	M\'{e}lanie~Chevance,\textsuperscript{1}\newauthor
	J.~M.~Diederik Kruijssen,\textsuperscript{1}\thanks{E-mail: kruijssen@uni-heidelberg.de}
	Mark~R.~Krumholz,\textsuperscript{2,4,5,6}
	and Steven~N.~Longmore\textsuperscript{7}
	\\
	\vspace{-1mm}\textsuperscript{1}Astronomisches Rechen-Institut, Zentrum f{\"u}r Astronomie der Universit{\"a}t Heidelberg, M{\"o}nchhofstra{\ss}e 12--14, 69120 Heidelberg, Germany\\
	\vspace{-1mm}\textsuperscript{2}Research School of Astronomy and Astrophysics, Australian National University, Canberra 2611, A.C.T., Australia\\
	\vspace{-1mm}\textsuperscript{3}Earth and Planets Laboratory, Carnegie Institution for Science, 5241 Broad Branch Road, NW, Washington, DC 20015, USA\\
	\vspace{-1mm}\textsuperscript{4}Centre of Excellence for Astronomy in Three Dimensions (ASTRO-3D), Australia\\
	\vspace{-1mm}\textsuperscript{5}Institut f\"{u}r Theoretische Astrophysik, Zentrum f{\"u}r Astronomie der Universit{\"a}t Heidelberg, Albert-\"{U}berle-Stra\ss e 2, 69120 Heidelberg, Germany\\
	\vspace{-1mm}\textsuperscript{6}Max-Planck Institut f\"{u}r Astronomie, K\"{o}nigstuhl 17, 69117 Heidelberg, Germany\\
	\vspace{-1mm}\textsuperscript{7}Astrophysics Research Institute, Liverpool John Moores University, IC2, Liverpool Science Park, 146 Brownlow Hill, Liverpool L3 5RF, United Kingdom\vspace{-3mm}
}

\date{\vspace{-4mm}Accepted 2020 July 21. Received 2020 July 21; in original form 2019 August 23}

\pubyear{2020}

\begin{document}
	\label{firstpage}
	\pagerange{\pageref{firstpage}--\pageref{lastpage}}
	\maketitle

	\begin{abstract} % 250 words
		Recent observational studies aiming to quantify the molecular cloud lifecycle require the use of known \enquote{reference time-scales} to turn the relative durations of different phases of the star formation process into absolute time-scales. We previously constrained the characteristic emission time-scales of different star formation rate (SFR) tracers, as a function of the SFR surface density and metallicity. However, we omitted the effects of dust extinction. Here, we extend our suite of SFR tracer emission time-scales by accounting for extinction, using synthetic emission maps of a high-resolution hydrodynamical simulation of an isolated, Milky-Way-like disc galaxy. The stellar feedback included in the simulation is inefficient compared to observations, implying that it represents a limiting case in which the duration of embedded star formation (and the corresponding effect of extinction) is overestimated. Across our experiments, we find that extinction mostly decreases the SFR tracer emission time-scale, changing the time-scales by factors of 0.04--1.74, depending on the gas column density. UV filters are more strongly affected than H${\alpha}$ filters. We provide the limiting correction factors as a function of the gas column density and flux sensitivity limit for a wide variety of SFR tracers. Applying these factors to observational characterisations of the molecular cloud lifecycle produces changes that broadly fall within the quoted uncertainties, except at high kpc-scale gas surface densities ($\Sigma_{\rm g}\ga20~{\mathrm{M_{\odot}\,pc^{-2}}}$). Under those conditions, correcting for extinction may decrease the measured molecular cloud lifetimes and feedback time-scales, which further strengthens previous conclusions that molecular clouds live for a dynamical time and are dispersed by early, pre-supernova feedback.
	\end{abstract}

	\begin{keywords}
		galaxies: evolution -- galaxies: ISM -- galaxies: star formation -- galaxies: stellar content -- dust, extinction -- H\,\textsc{ii} regions\vspace{-4mm}
	\end{keywords}

	%%%%%%%%%%%%%%%%%%%%%%%%%%%%%%%%%%%%%%%%%%%%%%%%%%

	%%%%%%%%%%%%%%%%% BODY OF PAPER %%%%%%%%%%%%%%%%%%
	\section{Introduction}
	One of the fundamental challenges of studying star formation is the assignment of time-scales to the underlying physical processes.
	The gas clouds within which stars form do not possess easily-observable natural clocks that can be used to estimate their histories and lifetimes.
	The stellar populations that these clouds form, however, do provide a natural clock dictated by the physics of stellar evolution.
	In particular, only the most massive and short-lived stars produce significant quantities of far- and extreme-ultraviolet (FUV and EUV) radiation, and thus observations of such emission provides direct constraints on the ages of the stars producing them \citep[e.g.][]{HAO11,KENN12}.
	In a galaxy where one observes both gas and stars, the statistics of the spatial correlation between young massive stars and gas can then be used to deduce time-scales for gaseous process for which the time-scales could not otherwise be constrained.
	The earliest uses of this technique relied on the presence of a large-scale spiral pattern \citep{TAMB08} or required the construction of catalogues of clouds and star clusters \citep{KAWA09}.
	However, more recently \citet{KRUI14} and \citet{KRUI18} refined this method (under the name \enquote{uncertainty principle for star formation}) into a general statistical tool that constrains gas evolutionary time-scales without requiring either the presence of a large-scale pattern or the construction of catalogues of clouds and star clusters.
	Instead, the method operates on arbitrary maps of tracers of gas and star formation.
	The development of the \textsc{Heisenberg} code based on this method \citep{KRUI18} has made it possible to empirically characterise the molecular cloud lifecycle across a dozen nearby galaxies \citep{KRUI19,CHEV20,WARD20,ZABE20}.

	While the \citet{KRUI14} method does not require catalogue construction, it does require calibration.
	By itself, the statistics of the correlation between tracers of gas and star formation constrain only relative time-scales between them.
	To assign absolute time-scales to the gas, one requires knowledge of the time-scales associated with stellar emission in whatever band or bands are to be used; we refer to these time-scales as \enquote{reference time-scales}.
	Unfortunately, the reference time-scale for a given tracer is not a function of stellar evolution alone: the spatial distribution of the stars as they disperse from their birth sites, and stochastic sampling of the initial mass function in regions of low stellar mass, matters as well.
	Thus accurate calibrations must be derived from simulations.
	In \citet{HAYD18} we performed this exercise to obtain reference time-scales at a range of metallicities (since metallicity affects the stellar emission spectrum) for several tracers of recent star formation: 12 ultraviolet (UV) filters (from GALEX, Swift, and HST) covering a wavelength range 150--350~nm, as well as H${\alpha}$ with and without continuum subtraction.
	This calibration forms the foundation for the absolute time-scales in nearby galaxies derived by \citet{CHEV20} and others.

	However, in \citet{HAYD18} we omitted the effects of extinction on SFR tracer lifetimes.
	This was partially for the sake of convenience, but also because the effects of extinction can, in most cases, be significantly reduced if not completely corrected for \cite[e.g.][]{JAME05}.
	However, correcting for extinction is not always possible or practical.
	In particular, for many galaxies we have access to only a single H${\alpha}$ image or a single UV band, and thus common extinction correction methods such as the Balmer decrement \citep{BERM36} or the UV spectral slope \citep{CALZ94} may not be available.
	For this reason, it is helpful to have an alternative approach available, which we investigate in this paper.
	The central idea of our work is that extinction does not fundamentally change the underlying emission lifetime, but it does act to reduce the amount of observed emission; this results in a different (effective) emission lifetime.
	We can calibrate the time-scales associated with extincted tracers using much the same approach that we used in \citet{HAYD18} to calibrate the unextincted ones.
	This allows us to make estimates for gas evolution time-scales from gas maps combined with maps of star formation tracers, even when we are not able to make an explicit extinction correction to the tracer maps.

	The structure of this paper is as follows.
	In \cref{sec:method}, we summarise our statistical analysis method applied through the \textsc{Heisenberg} code, describe the details of the simulations, and the procedure followed to generate synthetic emission maps.
	In \cref{sec:extinction}, we investigate and discuss how reference time-scales are affected as a result of extinction and in \cref{sec:sensitivity}, how this can be altered further by changing sensitivity limits.
	We summarise our findings in \cref{sec:conclusion}.

	\section{Method}\label{sec:method}

	\begin{table}
		\centering
		\caption{
			The star formation rate tracers we consider in this paper.
		}\label{tab:filterList}
		\subfloat[The UV filters we consider.
${\overline{\lambda}_{\mathrm{w}}}$ is the response-weighted mean wavelength of the filter.\label{tab:UVfilterList}]{%
			\begin{minipage}{\columnwidth}
				\centering
				\begin{tabular}{lllc}
					\toprule
					Telescope & Instrument & Filter & ${\overline{\lambda}_{\mathrm{w}} \left[\mathrm{nm}\right]}$\tabularnewline
					\midrule
					GALEX & & FUV & 153.9\tabularnewline
					GALEX & & NUV & 231.6\tabularnewline
					Swift & UVOT & M2 & 225.6\tabularnewline
					Swift & UVOT & W1 & 261.7\tabularnewline
					Swift & UVOT & W2 & 208.4\tabularnewline
					HST & WFC3 & UVIS1 F218W & 223.3\tabularnewline
					HST & WFC3 & UVIS1 F225W & 238.0\tabularnewline
					HST & WFC3 & UVIS1 F275W & 271.5\tabularnewline
					HST & WFC3 & UVIS1 F336W & 335.8\tabularnewline
					HST & WFPC2 & F255W & 259.5\tabularnewline
					HST & WFPC2 & F200W & 297.4\tabularnewline
					HST & WFPC2 & F336W & 335.0\tabularnewline
					\bottomrule
				\end{tabular}
			\end{minipage}
		}\\%
		\subfloat[The H${\alpha}$ filters we consider.\label{tab:HAfilterList}]{%
			\begin{minipage}{\columnwidth}
				\centering
				\begin{tabular}{p{0.15\columnwidth} p{0.75\columnwidth}}
					\toprule
					Filter & Details\tabularnewline
					\midrule
					H${\alpha{-}}$ & H${\alpha}$ emission with continuum subtraction.
                    This is calculated directly from the hydrogen-ionizing photon emission, see \cref{sec:emiMap} for details.\tabularnewline
					${\mathrm{H\alpha{+}}\,W}$ & A narrow band filter contain both H${\alpha}$ and the continuum.
                    The filter is defined in \cref{eq:HAFilter}.
					The total filter width is indicated by ${W}$; we consider ${W = \left\{10,~20,~40,~80,~160\right\}~\text{\AA}}$.
					\tabularnewline
					\bottomrule
				\end{tabular}
			\end{minipage}
		}
	\end{table}

	In \citet{HAYD18}, we presented measurements of the emission lifetimes for 18 SFR tracer filters.
	These measurements were, however, based on extinction-free emission maps, giving us the extinction-free emission lifetime, ${t_{\NoExt{}}}$.
	In this paper we want to understand how extinction can alter the measured emission lifetime.
	Here we outline the method we use to constrain \relDur{}: the factor by which the time-scale for an SFR tracer emission map without extinction (${t_{\NoExt{}}}$) changes when including extinction (${t_{\Ext{}}}$).
	We state this more explicitly:
	\begin{equation}\label{eq:relDur}
		\relDur{} = \frac{t_{\Ext{}}}{t_{\NoExt{}}}~\mathrm{.}
	\end{equation}
	We constrain \relDur{} for the same 18 SFR filters we previously considered and summarise these filters here in \cref{tab:filterList}. This list of filters includes several UV wavelengths and continuum-subtracted H$\alpha$, as well as a set of narrow band H$\alpha$ filters with different filter widths.\footnote{In observational applications of \textsc{Heisenberg}, filters of width $W>20$~\AA\ include contamination from other emission lines such as [N{\sc ii}] and [S{\sc ii}]. The results of this paper are not affected by such contamination, because the impact of dust extinction on these lines is very similar to that of H$\alpha$ due to their close proximity in wavelength. In a broader context, \citet{HAYD18} presents the emission time-scales for the same filters without accounting for extinction. While contamination from other lines could affect these calibrations at some level, [N{\sc ii}] and [S{\sc ii}] are emitted by H{\sc ii} regions, just like H$\alpha$, so we expect their emission time-scales to be similar. Finally, contamination is potentially important when inferring absolute SFRs from the H$\alpha$ flux. In that context, observational applications of this method generally adopt a constant correction factor throughout the galaxy to subtract the contamination \citep[e.g.][]{KRUI19,CHEV20}. With facilities like MUSE, it is now becoming possible to assess the environmental dependence of this correction factor \citep[e.g.][]{KREC16}.}

	In \cref{sec:KL14}, we give an overview of the \textsc{Heisenberg} code, which we use to measure \relDur{} directly.
	Our measurements of \relDur{} are based on the simulated galaxy we describe in \cref{sec:galSim}.
	From this simulated galaxy, we produce two groups of synthetic emission maps: emission maps with extinction and emission maps without (detailed in \cref{sec:emiMap}).
	The pair of synthetic SFR tracer maps (i.e.\ extincted and unextincted) are used as input for \textsc{Heisenberg}.

	\subsection{Analysis framework}\label{sec:KL14}
	The analysis we perform here makes use of the \textsc{Heisenberg} code.
	Since the code plays an important role in measuring \relDur{}, we give a summary here of the procedure used by \textsc{Heisenberg} and refer the reader to \cite{KRUI18} for the full details.

	As input, \textsc{Heisenberg} takes two emission maps that capture two phases of an evolutionary process (and their possible temporal overlap).
	For example, the star formation process could be captured by a CO emission map (tracing dense gas from which stars form) and an H${\alpha}$ emission map (tracing young massive stars).
	The three free parameters returned by \textsc{Heisenberg} would be the typical separation length between independent star forming regions, and the relative duration of the gas and overlap phases to the stellar phase (i.e.\ ${t_{\mathrm{gas}} / t_{\mathrm{star}}}$ and ${t_{\mathrm{over}} / t_{\mathrm{star}}}$, respectively).
	These durations would be converted into absolute values (i.e.\ ${t_{\mathrm{gas}}}$ and ${t_{\mathrm{over}}}$, respectively) if the value of ${t_{\mathrm{star}}}$ is known a priori.
	Determining ${t_{\mathrm{star}}}$ is the aim of the work we present here and also in \citet{HAYD18}.

	\textsc{Heisenberg} constrains these parameters by first identifying all the emission peaks present in both images within the range ${\left[10^{\left(E^{\max}_{i}  - \Delta\log_{10}\mathcal{F}_{i}\right)},~10^{E^{\max}_{i}}\right]}$, where ${i = \left\{\mathrm{star},~\mathrm{gas}\right\}}$, ${E^{\max}_{i}}$ is the base-ten logarithm of the maximum emission in map ${i}$, and ${\Delta\log_{10}\mathcal{F}_{i}}$ defines the depth of the range in map ${i}$ (a parameter specified by the user in the input file).
	This depth (approximately) corresponds to the dynamic range of the map, from the brightest emission peak to the noise floor.
	We place apertures centred on these emission peaks.
	The total flux across all apertures identified in the stellar, and then gas, map is calculated for both emission maps.
	This gives four measurements, ${T^{p}_{e}}$: the total gas/stellar emission (${e = \left\{\mathrm{star},~\mathrm{gas}\right\}}$) at the location of gas/stellar peaks (${p = \left\{\mathrm{star},~\mathrm{gas}\right\}}$).
	The ratios ${T^{p}_{\mathrm{gas}} /  T^{p}_{\mathrm{star}}}$, in units of the galactic average flux ratio, are calculated as a function of aperture size.
	The best fitting model to the two curves (i.e.\ ${p = \mathrm{star}}$ and ${p = \mathrm{gas}}$), gives the values of the three free parameters.

	When using \textsc{Heisenberg} we use the default input parameters that are listed in \citet[Tables~1 and~2]{KRUI18}; with the following exceptions.
	The set-up of the galaxy simulation necessitates the use of cuts in galactocentric radius: we use ${R_{\min} = 3}$~kpc and  ${R_{\max} = 11}$~kpc (and set the \texttt{cut\_radius} flag to 1).
	We also change the range of aperture sizes: we use a minimum aperture size of ${l_{\mathrm{ap,}\,\min} = 25}$~pc and have ${N_{\mathrm{ap}} = 17}$ apertures.
	This results in 17 logarithmically spaced aperture diameters from 25--6400~pc
	We set parameters which allows use to measure \relDur{} directly (\texttt{tstariso}~=~1, \texttt{tstar\_incl}~=~1).
	We also use ${\Delta\log_{10}\mathcal{F}_{\mathrm{star}} = \Delta\log_{10}\mathcal{F}_{\mathrm{gas}} \equiv \Delta = 2}$.\footnote{
		This is the default value, as given in \citet[Table~2]{KRUI18}; however, ${\Delta}$ plays a role in \cref{sec:extinction,sec:sensitivity} and so we explicitly state it here.
	}
	While ${\Delta = 2}$~dex is often appropriate for observational applications \citep[e.g.][]{KRUI19,CHEV20}, as well as for the simulation here, we would like to emphasis that this is not always the case.

	For this paper, we make \enquote{notational transformations} from the example used above (and the names of parameters given in \citealt{KRUI18}): \enquote{gas}~${\rightarrow}$~\enquote{\raisebox{1pt}{\Ext{}}} (emission map \emph{with} extinction) and \enquote{star}~${\rightarrow}$~\enquote{\raisebox{1pt}{\NoExt{}}} (emission map \emph{without} extinction).
	It can be seen after this transformation that one of the free quantities of the model, namely ${t_{\mathrm{gas}} / t_{\mathrm{star}}}$, is in fact \relDur{}.

	\subsection{Galaxy simulation}\label{sec:galSim}
	The work presented in this paper uses the high-resolution hydrodynamical simulation of an isolated Milky-Way-like disc galaxy described in \citet{FUJI18}.\footnote{This simulation differs from the one used in \citet[originally presented in \citealt{KRUI18}]{HAYD18}, because the lower resolution of that simulation prohibits estimating the dust column density towards individual H{\sc ii} regions. Thanks to the $\sim8$~pc resolution of the simulation used in the present paper, this is not a concern here.}
	We use the adaptive mesh refinement code \textsc{enzo} \citep{BRYA14} to simulate a 128~kpc box with cell sizes ranging from 7.8125~pc to 31.25~pc over seven levels of refinement during the initial stages of the simulation.
	The initial conditions of the simulation match those of \citet{TASK09}.
	We have a gas disc in a static background potential which accounts for both dark matter and the stellar disc.
	The axisymmetric background potential has a logarithmic form with a constant circular velocity of 200~${\mathrm{km\,s^{-1}}}$ at large galactocentric radii (${r > 2\ \mathrm{kpc}}$).
	The gas disc has an initial mass of ${8.6\times10^9~\mathrm{M_{\odot}}}$ and a density profile divided into three regions defined by a constant value of the Toomre ${Q}$ parameter \citep{TOOM64}: ${Q = 1}$ between galactic radii of 2--13~kpc, ${Q = 20}$  for 0--2~kpc and 13--14~kpc, and beyond 14~kpc is a static, very low density medium.
	The stellar bulge is not modelled and its effects are included implicitly through the potential.
	The simulation initially evolves over 730~Myr, wherein the allowed maximum resolution is gradually increased up to the highest resolution of 7.5~pc, and the galaxy settles into an equilibrium state (see \citealt{FUJI18} for details of our refinement strategy).
	We then evolve the simulation further; for the analysis here, we take the 850~Myr snapshot.

	Star particles, always with an initial mass of 300~${\mathrm{M_{\odot}}}$, form probabilistically according to a gas density threshold; however, this is only permitted between galactocentric radii of 2--14~kpc.
	Each star particle represents a simple stellar population, which we model using version 2 of the \textsc{slug} stochastic stellar population synthesis code \citep{SILV12, SILV14, KRUM15}.
	The stellar population associated to the star particle is created by sampling a \citet{CHAB05} IMF continuously up to a mass of 9~${\mathrm{M_{\odot}}}$ and then stochastically for more massive stars using the \texttt{POISSON} sampling method described in Appendix A of \citet{KRUM15}.
	The stellar population evolves during the simulation in accordance with the Padova stellar evolution tracks at solar metallicity \citep{GIRA00}.
	Stellar atmospheres are calculated with \textsc{Starburst99} \citep{LEIT99,VAZQ05} spectral synthesis models.
	The stellar populations associated to the star particles are important for the implemented stellar feedback models, as they determine the ionising luminosity (for photoionisation calculations) as well as the timing of supernova explosions.
	In order to conserve memory, these stellar populations only remain with the star particle up to the stellar age of 40~Myr.

	In \citet{FUJI19}, we show that the simulation is consistent with most observational constraints on scales ${\gtrsim 100}$~pc (e.g.\ the simulation reproduces the global- and kpc-scale Kennicutt-Schmidt relations for both molecular and total gas; and matches the observed decomposition of the interstellar medium into warm neutral, cold neutral, and molecular phases).
	However, on smaller (${\lesssim 100}$~pc) scales, it can be seen that the implemented feedback mechanisms are insufficient to disperse the surrounding gas.
	When comparing the synthetic SFR tracer (H${\alpha}$) and dense gas (CO ${J = 1\rightarrow0}$) emission maps of the simulated galaxy, it is clear that most (if not all) H${\alpha}$ emitting regions are co-spatial with CO emission.
	Observationally, this would result in a galaxy with a high amount of extinction and a long (overlap) phase where H${\alpha}$ and CO are co-spatial.
	This simulation therefore gives us the opportunity to constrain the characteristic time-scales of SFR tracers in a system with a high amount of extinction; this complements the work from \citet{HAYD18}, which excluded the effects of extinction.
	This also means that the results we find here should be considered as bounding limits.
	We expect that real observations would not be subjected to this level of extinction as typically the stellar feedback disrupts the parent cloud in a few Myr (as suggested by the decorrelation of H${\alpha}$ and CO on the cloud scale, see \citealt{KRUI19,CHEV20}).
	Therefore, real observations would have \relDur{} values that fall somewhere within the range spanned by the results presented here and ${\relDur{} = 1}$.

	\subsection{Generation of the emission maps}\label{sec:emiMap}
	We describe here how we create the synthetic SFR tracer emission maps of the simulated galaxy.
	In principle, the synthetic emission maps could be created using the simple stellar population associated to each of the star particles within the simulation.
	For two reasons, we instead choose to run the \textsc{slug2} model as a post-production process.
	Firstly, for consistency with \citet{HAYD18}, we wish to use the Geneva solar-metallicity evolutionary tracks \citep{SCHA92} instead of the Padova tracks used within the simulation.
	Secondly, as described in \cref{sec:galSim}, only star particles up to 40~Myr have an associated stellar population; this would impose an age cut in the emission maps at 40~Myr.
	The artificial age cut in the emission map would have little impact on the ${\mathrm{H\alpha}{\pm}}$ filters since their characteristic time-scales are short; however, for the WFC3 UVIS1 F336W filter, with ${t_{\mathrm{E,0}} = 33.3~\mathrm{Myr}}$ \citep{HAYD18}, this age cut could be problematic.

	For the post-production \textsc{slug2} simulations, we use Geneva solar-metallicity evolutionary tracks \citep{SCHA92} and \textsc{Starburst99} spectral synthesis.
	For the emission spectrum, we also include the contributions of the surrounding nebular material, which has a hydrogen number density of ${10^{2}~\mathrm{cm}^{-3}}$ and reprocesses 73~per~cent of the ionising photons into nebular emission.
	The 27~per~cent \enquote{loss} of ionising photons (either absorbed by circumstellar dust, or scattered outside the observational aperture) is consistent with the estimate from \citet{MCKE97}.
	When calculating the extinction, we use the Milky Way extinction curve that comes included with \textsc{slug2}.

	For consistency, we use the same population that was generated in the simulation, instead of creating a new stellar population for each star particle.
	This is achieved by extracting the stellar population from the snapshot file in which the star particle first appears (this is always before the stellar population reaches the age of 1~kyr).
	We do this for all the star particles present in the 850~Myr snapshot which first appear in or after the 730~Myr snapshot.
	This will introduce a new 120~Myr age cut into the emission map; however, this is sufficiently long as to not impact the results we present here.

	We assign to each star particle a visual extinction (in magnitudes), ${A_{V}}$, which \textsc{slug2} uses to compute an extincted emission spectrum.
	To calculate ${A_{V}}$ for each star particle we use the ray tracing feature in the python library \textsc{yt} \citep{TURK11}.
	The \enquote{ray} passes between the star particle located at (${x_{\star},~y_{\star},~z_{\star} }$) and the edge of the simulated box (${x_{\star},~y_{\star},~z_{\mathrm{edge}} }$); that is, we are looking at the galaxy face on.
	For a given cell in the simulation that the ray passes through, ${i}$, we recover the gas mass density, ${\rho_{\mathrm{g, i}}}$, as well as the distance travelled through the cell, ${d_{\mathrm{i}}}$.
	From this information we calculate the H nuclei number column density, ${N_{\mathrm{H}}}$, as
	\begin{equation}\label{eq:ColNumDen}
		N_{\mathrm{H}} = \frac{f_{\mathrm{H}} \sum_{i}\rho_{\mathrm{g, i}}d_{\mathrm{i}}}{m_{\mathrm{H}}}~\mathrm{,}
	\end{equation}
	where ${m_{\mathrm{H}}}$ is the mass of a H nuclei, and ${f_{\mathrm{H}} =  0.76}$ is the primordial H mass fraction.
	We calculate ${A_{V}}$ from ${N_{\mathrm{H}}}$ as \citep{SAFR17}
	\begin{equation}\label{eq:Av}
	A_{V} = N_{\mathrm{H}} \sigma_{\mathrm{d, V}}~\mathrm{,}
	\end{equation}
	where ${\sigma_{\mathrm{d, V}} = 5.3 \times 10^{-22}~\mathrm{cm^{2}}}$ \citep{DRAI96} is the total dust cross-section in the \textit{V}-band per H nucleus.

	The full combined photometric spectrum of the stellar population (with and without extinction) is passed through the response profile of the filters of interest (see \cref{tab:filterList}) to recover a single luminosity for each of filters; this value is what we assign to the star particle.
	The UV response filters used here are included by default in \textsc{slug2} (see \citealt{KRUM15} for more details).
	For H${\alpha{-}}$, we convert the hydrogen-ionizing photon emission into a true H${\alpha}$ luminosity using \citet[Equation~2]{SILV14}.
	For ${\mathrm{H\alpha{+}}\,W}$, we define the narrow band filter, ${\mathcal{F}_{\mathrm{H\alpha{+}}\,W}}$, as
	\begin{align}\label{eq:HAFilter}
		\mathcal{F}_{\mathrm{H\alpha{+}}\,W} = \left\lbrace
		\begin{aligned}
			&1 &&6562 -\frac{W}{2}~\text{\AA} \leq \lambda \leq 6562 +\frac{W}{2}~\text{\AA}
			\\
			&0 &&\mathrm{Otherwise}
		\end{aligned}
		\right.~\mathrm{.}
	\end{align}

	\begin{figure}
		\centering
		\includegraphics[width=\columnwidth]{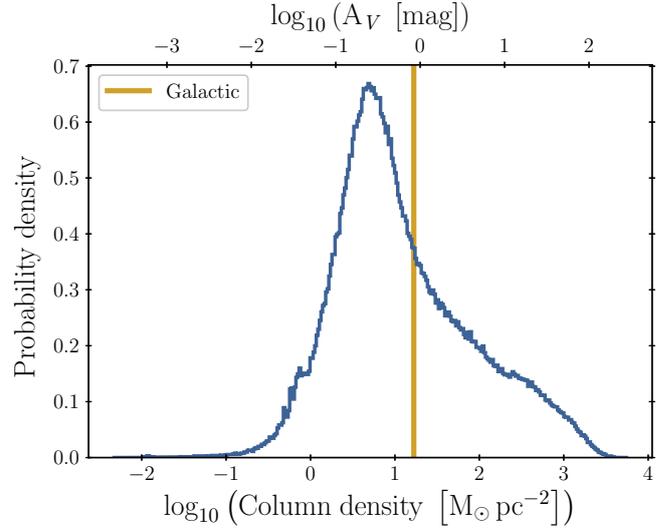}%
		\caption{
			A histogram showing the distribution of column densities and visual extinctions (calculated using \cref{eq:ColNumDen,eq:Av}) associated to the star particles used in creating the synthetic emission maps.
			The vertical yellow line indicates the galactic average column density (16.7~${\mathrm{M_{\odot}\,pc^{-2}}}$) / visual extinction (0.84~mag).
			This is calculated from the average gas surface density between the galactocentric radii 3--11~kpc (as marked in \cref{fig:HALUM}).
		}\label{fig:ColDen}
	\end{figure}

	In \cref{fig:ColDen} we show the distribution of column densities (and visual extinction) that is associated to the star particles used in creating the synthetic emission maps (i.e.\ stars surviving the 120~Myr age cut).
	In the figure, we also mark a \enquote{galactic column density} found to be 16.7~${\mathrm{M_{\odot}\,pc^{-2}}}$ (corresponding to ${A_{V}=0.84~\mathrm{mag}}$).
	This is calculated from the average gas surface density of the 850~Myr snapshot (without an age cut), ${\overline{\Sigma}_{\mathrm{g}}}$, between the galactocentric radii 3--11~kpc by using \cref{eq:ColNumDen} with ${\overline{\Sigma}_{\mathrm{g}}}$ in place of the summation term.

	\begin{figure*}
		\centering
		\includegraphics[width=\textwidth]{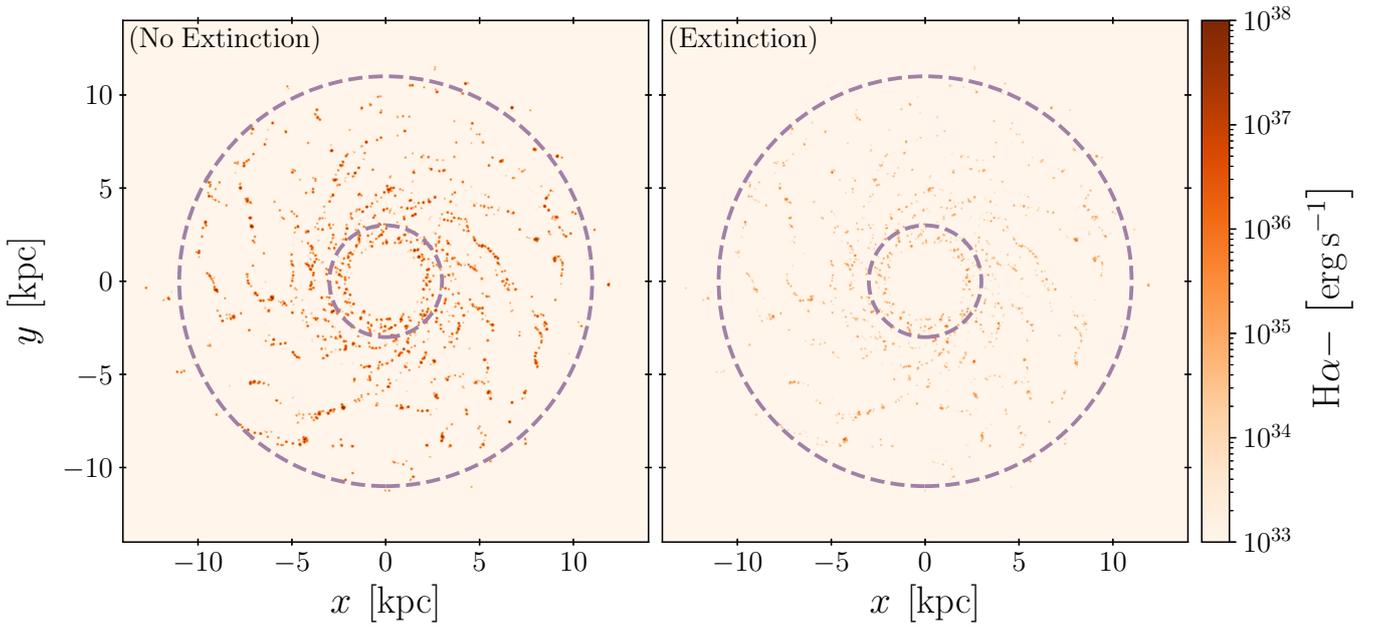}%
		\caption{
			A comparison between synthetic emission maps without (left) and with (right) extinction for continuum subtracted H${\alpha}$ emission (H${\alpha{-}}$).
			The purple dashed annuli, with inner radius of 3~kpc and outer radius of 11~kpc, indicate the region of the galaxy we use for our analysis.
		}\label{fig:HALUM}
	\end{figure*}

	In \cref{fig:HALUM}, we present example synthetic H${\alpha{-}}$ emission maps of the simulated galaxy, which are used in our analysis.
	The two presented maps allow us to see how extinction reduces the amount of emission present in the image.
	The total emission in the extincted map is 2.1~per~cent of the emission in the unextincted map.
	The figure also indicates the region of the galaxy we will use in our analysis with the \textsc{Heisenberg} code.
	We exclude the inner 3~kpc and beyond 11~kpc of the galaxy due to the way in which the galaxy simulation was set-up (see \cref{sec:galSim}).

	The two groups of emission maps we produce here are related to each other.
	Excluding the effects of extinction, they are in fact identical; that is, all emitting regions are present in both maps.
	This violates the first requirement listed in \citet[][Section~4.4]{KRUI18}, which details the conditions under which \textsc{Heisenberg} should be used.
	The \textsc{Heisenberg} code assumes that the pair of maps analysed traces two different phases of an underlying evolutionary process.
	With this assumption, it is expected that within the galaxy many emission peaks in one emission map will be independent of the other and only some regions will be present in both maps (the \enquote{overlap} phase).
	We circumvent this aspect of the method by spatially transforming (through rotation and/or reflection) one emission map relative to the other.
	This transformation will spatially offset the emitting regions and give the impression of an evolutionary sequence.
	In performing the transformation, we forfeit information about two quantities constrained by \textsc{Heisenberg} (the spatial separation of regions and the duration of the overlap phase) whilst retaining the ability to measure the relative duration of one phase to the other: this is unaffected by the relative positions of regions (see the discussion in \citealt{KRUI14} and \citealt{KRUI18}).

	Relying on the fact that the relative duration of an emission map to itself is always unity, we can test to find which transformation works best given the specific structure and morphology of the adopted galaxy simulation.
	Specifically, we test for which transformation \textsc{Heisenberg} returns a relative duration consistent with unity and a good model fit, with reduced-${\chi^{2} \approx 1}$, when applied to a given map and the spatially transformation version of itself.
	We consider either no reflection \enquote{F0} or a left-right reflection \enquote{F1} and four rotations ${\left\lbrace 0\degree,~90\degree,~180\degree,~270\degree\right\rbrace}$ denoted as \enquote{R${X}$}, where ${X}$ is the angle of rotation.
	This gives the seven possible unique transformations (excluding no transformation, \enquote{F0 R0}).

	\begin{figure}
		\centering
		\includegraphics[width=\columnwidth]{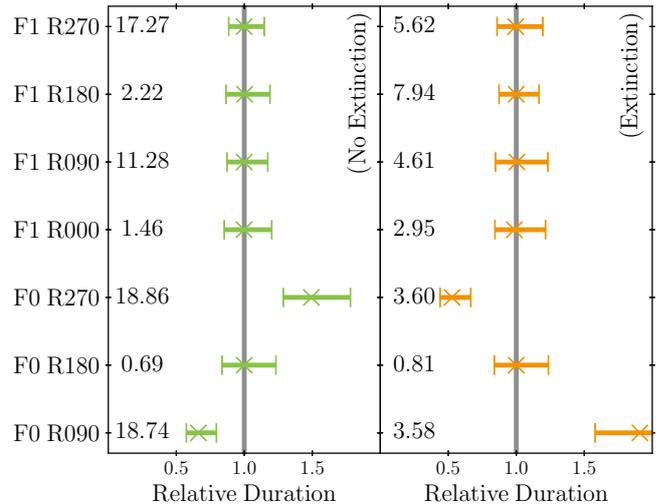}%
		\caption{
			Results showing the measured relative duration of an emission map when compared to a spatially transformed version of itself.
			Left: H${\alpha{-}}$ emission map without extinction.
			Right: H${\alpha{-}}$ emission map with extinction.
			The y-axis labels, \enquote{F${X}$ R${Y}$}, indicate the spatial transformation applied to one of the two input maps: ${X}$ indicates if a left-right reflection was used, and ${Y}$ gives the angle of rotation.
			Note that \enquote{F0 R0} (no transformation) is excluded as it violates the conditions given in \citet[][Section~4.4]{KRUI18}.
			The relative duration of an emission map to itself should be unity and so we have marked this value with a vertical grey line.
			The reduced-${\chi^{2}}$ value of the model fit is also reported alongside each measurement of the relative duration.
			The optimal reduced-${\chi^{2}}$ value is unity.
			The error bars indicate the uncertainties on the measurements derived from the ${\chi^2}$ fit carried out by \textsc{Heisenberg}.
		}\label{fig:TranTest}
	\end{figure}

	\cref{fig:TranTest} shows the relative durations we measure using \textsc{Heisenberg} when applying each of the seven spatial transformations to the H${\alpha{-}}$ emission maps.
	For all of the SFR tracers we consider, we find that using a left-right reflection always gives the expected relative duration of unity.
	Of the \enquote{F1} series, the best reduced-${\chi^{2}}$ is for zero rotation.
	By contrast, a purely rotational transformation without reflection (\enquote{F0}) does not always return a relative duration of unity.
	This is likely caused by the symmetric nature of the spiral arms, which causes large parts of them to overlap at rotations of ${90\degree}$ and ${270\degree}$.
	In all the experiments that follow, we will always apply the \enquote{F1 R0} transformation to the extincted emission map.

	\section{Extinction}\label{sec:extinction}
	By using the \textsc{Heisenberg} code on two emission maps, we quantify the relative duration of one map to the other.
	For example, if we have a gas tracer map (e.g.\ CO) and an SFR tracer map (e.g.\ H${\alpha}$), we can measure the relative duration of the CO map to the H${\alpha}$ map.
	With the known characteristic time-scale of H${\alpha}$ emission, we can then recover an absolute time-scale for the CO emission.
	However, extinction reduces the overall emission within an emission map and possibly reduces the number of identifiable star-forming regions.
	The relative duration of the same CO emission map to an \emph{extincted} H${\alpha}$ map will differ and so should the characteristic time-scale we associate to the extincted H${\alpha}$ emission map.
	Here we quantify by how much the characteristic time-scales we associate to various SFR tracers differ when dealing with extincted emission.
	As previously mentioned, the galaxy we are using for this analysis has heavily extincted emission due to inefficient feedback \citep{FUJI19}: most, if not all, star forming regions are still embedded in gas.
	This means that the differences in characteristic time-scales we would expect for real galaxies would be less extreme than those we find here: the results here represent a limiting case, illustrating the maximum effect induced by extinction.

	We provide the \textsc{Heisenberg} code with an emission map and a spatially transformed (\enquote{F1 R0}) extincted emission map and recover \relDur{}, the relative duration of the extincted emission map compared to the unextincted map.
	This is the factor by which the characteristic emission time-scale of the SFR tracer changes as a result of extinction.
	In order to see how the characteristic time-scales change as a function of extinction we repeat the experiment with different extincted emission maps.
	We produce these different extincted emission maps by modifying \cref{eq:ColNumDen} to include a density scaling factor, ${F_{\mathrm{d}}}$, such that ${N_{\mathrm{H}}^{\prime} \equiv F_{\mathrm{d}}N_{\mathrm{H}}}$.
	This scaling factor multiplies the gas density used to calculate the ${A_{V}}$ associated to the star particle, which \textsc{slug2} subsequently uses to calculate the extincted emission.
	We choose scaling factors in the range ${\left[10^{-3},~10^{1}\right]}$ in 0.25 logarithmic steps.
	The native simulation column density is at ${F_{\mathrm{d}} = 1}$.

	In this work, we only consider solar metallicity.
	Any change in metallicity is expected to linearly translate into an analogous change of the dust column density, such that the metallicity change can be absorbed into the column density scaling factor, ${F_{\mathrm{d}}}$, defined here.
	As such, the dependence on column density quantified here also reflects the expected dependence on metallicity at fixed gas surface density.
	If a metallicity gradient is present with a total dynamic range larger than the uncertainty on the dust column density, it is necessary to divide the galaxy into bins of galactocentric radius and calculate $\relDur$ for each of these. A similar approach was taken in applications of the method to date when accounting for the metallicity dependence of the reference time-scale or the CO-to-H$_2$ conversion factor \citep[see e.g.][]{KRUI19,CHEV20}.

	Likewise, our analysis of the synthetic emission maps always assumes a face-on orientation.
	However, our results can still be extended to observations of galaxies that are not face on.
	As a galaxy is viewed under a higher inclination angle, the observed average gas column density increases.
	In order to correct for an inclination angle ${i}$, this means that the column density scaling factor ${F_{\mathrm{d}}}$ should be scaled by a factor of ${{\left[\cos\left(i\right)\right]}^{-1/2}}$.
	In doing so, we note that there are limits on the inclination angle for which \textsc{Heisenberg} still provides reliable measurements \citep[ideally, ${i \lesssim 70 \degree}$; for details see Section~4.3.7 of][]{KRUI18}.

	\begin{figure}
		\centering
		\includegraphics[width=\columnwidth]{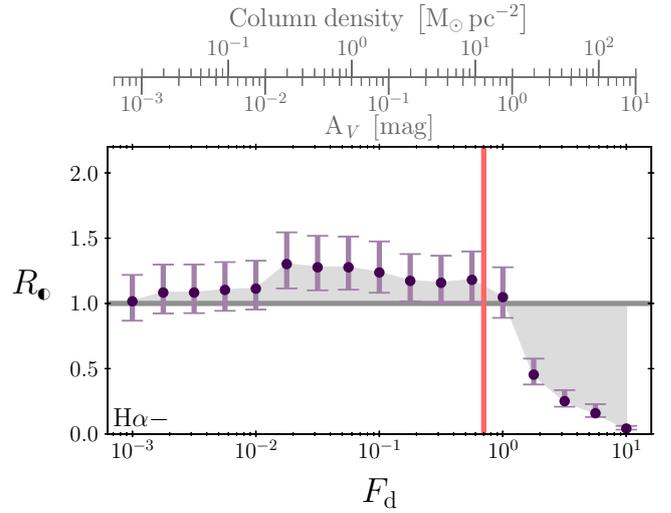}%
		\caption{
			Relative amount by which the continuum-subtracted H${\alpha}$ (H${\alpha{-}}$) emission time-scale changes due to extinction.
			Shown is the relative duration of the time-scale associated to the extincted compared to the unextincted SFR tracer emission map, \relDur{}, as a function of the column density scaling factor, ${F_{\mathrm{d}}}$.
			A scale factor of unity indicates column densities taken directly from the simulation.
			The horizontal grey line (at ${\relDur{} = 1}$) indicates the point where the time-scale associated to each of the two emission maps is the same.
			The vertical red line indicates the approximate location where we expect \relDur{} to transition from ${\relDur{} > 1}$ to ${\relDur{} < 1}$ based on the flux density distribution of the maps (see the text and \cref{fig:extDistHA-} for details).
			The top grey axis converts ${F_{\mathrm{d}}}$ into a column density and visual extinction ${A_{V}}$ based on a column density equal to the average gas surface density, ${\overline{\Sigma}_{\mathrm{g}} = 16.7~\mathrm{M_{\odot}\,pc^{-2}}}$.
			Since our experiments probe the maximum effects of extinction, we expect to find \relDur{} for real galaxies to lie within the grey-shaded region (i.e.\ between the unity line and the data points shown).
		}\label{fig:scaleHA-}
	\end{figure}
	\begin{figure}
		\centering
		\includegraphics[width=\columnwidth]{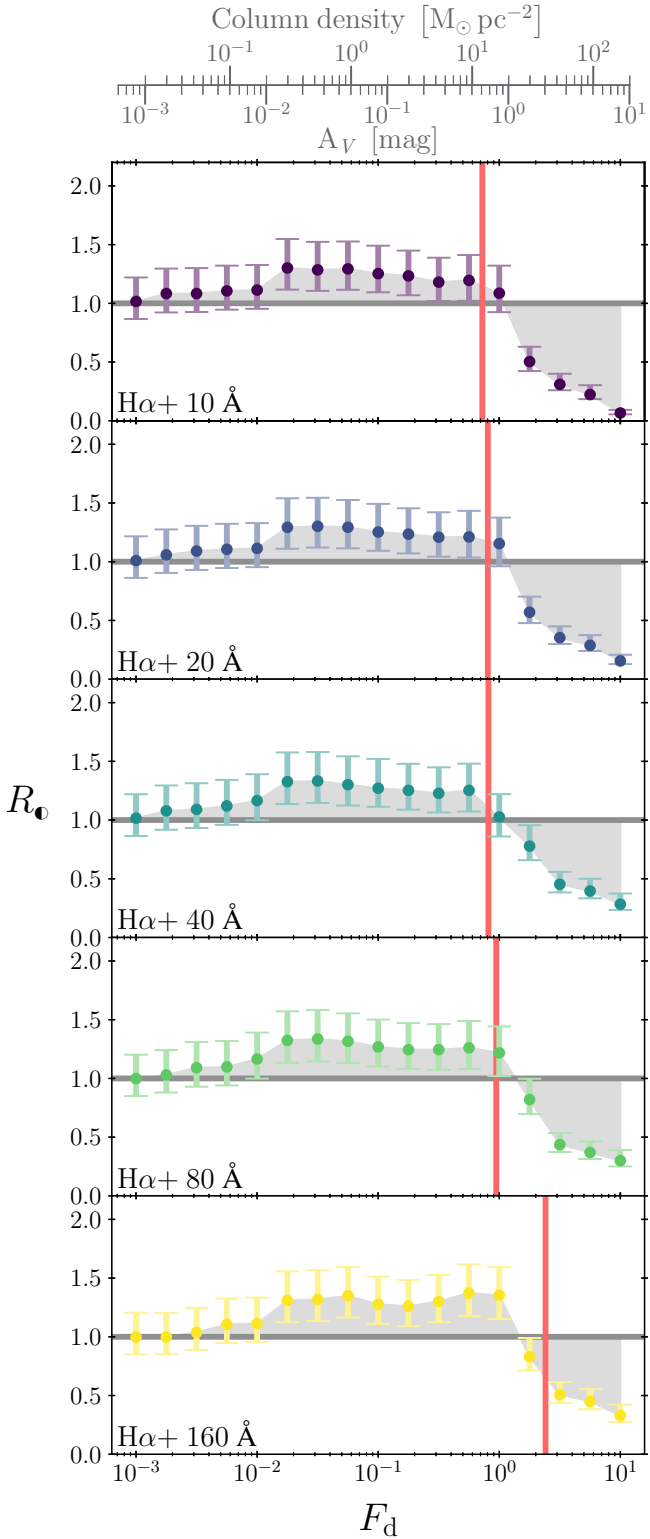}
		\caption{
			Same as \cref{fig:scaleHA-} for H${\alpha}$ filters without continuum subtraction (H${\alpha{+}}$).
			The filter width is indicated in the bottom left corner of each panel.
		}\label{fig:scaleHA+}
	\end{figure}
	\begin{figure*}
		\centering
		\includegraphics[width=0.9\textwidth]{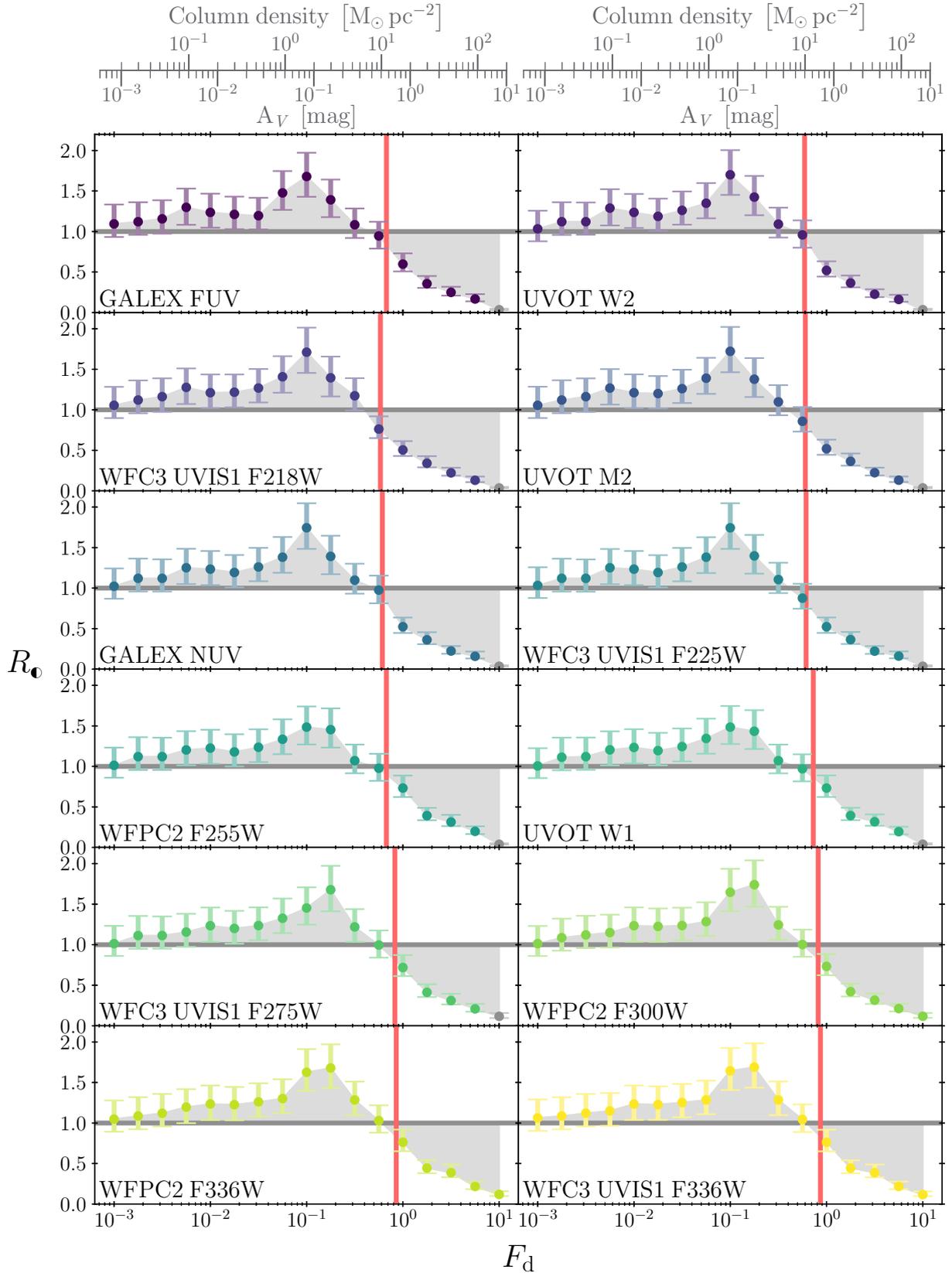}%
		\caption{
			Same as \cref{fig:scaleHA-} for UV emission filters.
			The panels are ordered from left to right, top to bottom, by the response-weighted mean wavelength.
			The filter name is indicated in the bottom left corner of each panel.
			The grey-shaded data points have fewer than 35 identified emission peaks in the extincted emission image and are therefore unreliable \citep[see][Section~4.4]{KRUI18}.
		}\label{fig:scaleUV}
	\end{figure*}

	In \cref{fig:scaleHA-,fig:scaleHA+,fig:scaleUV} we show how \relDur{} changes with density scaling factor, ${F_{\mathrm{d}}}$, for the H${\alpha{-}}$, H${\alpha{+}}$, and UV filters.
	Most importantly, we find that extinction does not affect the H${\alpha}$ tracer emission time-scales at gas surface densities ${\Sigma_{\mathrm{g}} \lesssim 20~\mathrm{M_{\odot}~{pc}^{-2}}}$.
	The UV tracer emission time-scales start to become affected at gas surface densities ${1~\mathrm{M_{\odot}~{pc}^{-2}} \lesssim\Sigma_{\mathrm{g}} \lesssim 10~\mathrm{M_{\odot}~{pc}^{-2}}}$.
	At the extremes of ${F_{\mathrm{d}}}$ (or equivalently, column density), the observed behaviour is as expected.
	As the column density tends towards zero, \relDur{} tends towards unity.
	This is because, as the column density decreases, the amount of extinction is reduced, which results in an extincted emission map more like the emission map without extinction.
	As the column density increases, the amount of extinction increases and \relDur{} tends towards zero as more emission is removed from the map.
	In practice, the results from \textsc{Heisenberg} become unreliable once fewer than 35 regions are identified and so \relDur{} will instead tend towards some small non-zero value.
	The grey-shaded data points in \cref{fig:scaleUV} (not present in \cref{fig:scaleHA-,fig:scaleHA+}) indicate these unreliable results, which are typically found for ${F_{\mathrm{d}} \gtrsim 10}$ (or ${\Sigma_{\mathrm{g}} \gtrsim 200~\mathrm{M_{\odot}~{pc}^{-2}}}$).

	\begin{figure*}
		\centering
		\includegraphics[width=\textwidth]{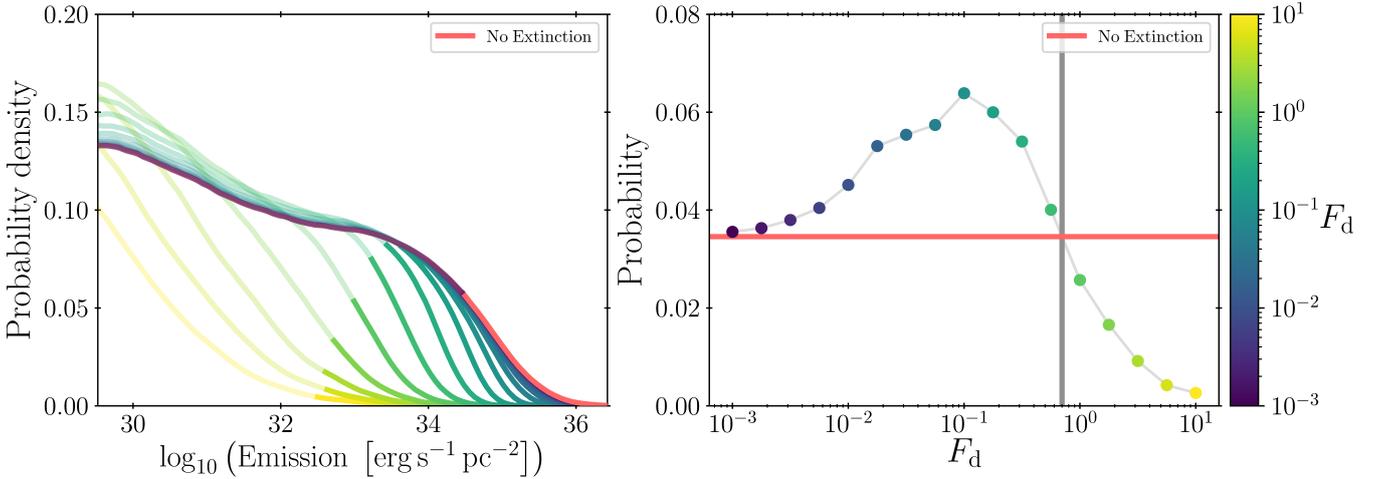}%
		\caption{
			The left-hand panel shows the high-emission end of the H${\alpha{-}}$ probability density distribution of emission in emission maps with different levels of extinction.
			The colour of the line indicates the density scaling factor, ${F_{\mathrm{d}}}$, used in creating the emission map.
			Higher values of ${F_{\mathrm{d}}}$ correspond to higher amount of extinction, the native amount of extinction occurs at ${F_{\mathrm{d}} = 1}$ (or ${\Sigma_{\mathrm{g}}=16.7~\mathrm{M_{\odot}~{pc}^{-2}}}$).
			The red line indicates the results for the emission map without extinction.
			The non-transparent line segments indicate the range of the distribution that is within ${\Delta = 2}$~dex (see \cref{sec:KL14} for details) of the maximum.
			The right-hand panel shows the probability (or fraction) of emission falling within ${\Delta}$ of the maximum (i.e.\ area under the non-transparent line segments in the left-hand panel).
			The vertical grey line indicates the approximate ${F_{\mathrm{d}}}$ value for the transition between ${P_{\Ext{}} > P_{\NoExt{}}}$ and ${P_{\Ext{}} < P_{\NoExt{}}}$, where ${P_{\Ext{}}}$ is the probability of finding emission within ${\Delta}$ of the maximum in the emission map with extinction and ${P_{\NoExt{}}}$ without (shown as the horizontal red line).
			We see that, at low levels of extinction (${F_{\mathrm{d}} \lesssim 1}$), a larger fraction of the emission resides in the top ${\Delta}$ of the distribution, as a result of only the brightest emission peaks being affected; that is, those peaks which formed at the highest gas column densities.
			By contrast, at high levels of extinction (${F_{\mathrm{d}} \gtrsim 1}$), all emission peaks are affected (resulting in the suppression of the \enquote{knee} in the distribution in the left-hand panel) and a smaller fraction of emission resides in the top ${\Delta}$ of the distribution.
		}\label{fig:extDistHA-}
	\end{figure*}

	One might expect that the value of \relDur{} should always be in the range [0, 1] (i.e.\ a shorter emission lifetime), as extinction removes emission from the image.
	However, the results in  \cref{fig:scaleHA-,fig:scaleHA+,fig:scaleUV} show that for ${F_{\mathrm{d}} \lesssim 1}$ (or ${\Sigma_{\mathrm{g}} \lesssim 10~\mathrm{M_{\odot}~{pc}^{-2}}}$), \relDur{} typically falls in the range [1, 1.5].
	We use the results in \cref{fig:extDistHA-} to explain this behaviour.
	In the left panel of \cref{fig:extDistHA-}, we show the upper end of the emission probability density distribution.
	The area under this curve gives the probability of any point in the emission map having a given amount of emission.
	We can see that as ${F_{\mathrm{d}}}$ increases the distribution moves towards the left, i.e.\ lower emission values become more likely.
	This shows that extinction removes emission from the map as a whole.
	The overall shift in the emission distribution has minimal effect on the results from \textsc{Heisenberg}: absolute emission values are not relevant to the measurement of \relDur{}.
	The \textsc{Heisenberg} code identifies emission peaks based on contours that start from the maximum emission value in the image and go down for a certain maximal range, ${\Delta = 2}$~dex (see \cref{sec:KL14} for details).
	This means that the fraction of the total emission contained in this interval ${\Delta}$ is what impacts \relDur{} the most.

	The non-transparent line segments in \cref{fig:extDistHA-} show the part of the distribution within ${\Delta}$ of the maximum emission value present in the emission map.
	It is clear that extinction alters the shape of the emission probability density distribution, but what is of particular note is the gradients of these ${\Delta}$ line segments.
	When extinction is initially introduced (small ${F_{\mathrm{d}}}$), the gradients of these segments become steeper than that without extinction.
	This shows that extinction has preferentially removed emission from the highest emission peaks, because the brightest emission peaks form at the highest gas column densities.
	Since the gradients are steeper, the area under the curve is larger (for the same change in emission, i.e.\ ${\Delta}$) and, as shown in the right-hand panel of \cref{fig:extDistHA-}, the probability of finding emission within ${\Delta}$ of the maximum for the extincted emission map (${P_{\Ext{}}}$) is higher than for the unextincted emission map (${P_{\NoExt{}}}$).
	In constraining \relDur{}, \textsc{Heisenberg} will then detect more emission peaks within ${\Delta}$ of the maximum compared to the unextincted image.
	As a result, the duration associated with the extincted emission map is likely to be longer, thus causing ${\relDur{} > 1}$.
	
	Using linear interpolation, we find the ${F_{\mathrm{d}}}$ that corresponds to ${P_{\Ext{}} = P_{\NoExt{}}}$ for each of the SFR tracers and mark this transition point in \cref{fig:scaleHA-,fig:scaleHA+,fig:scaleUV} (with a vertical red line).\footnote{The fact that this line does not always match the transition in the data points in \cref{fig:scaleHA-,fig:scaleHA+,fig:scaleUV} results from the stochasticity caused by using a simple linear interpolation of only two data points on either side of ${P_{\Ext{}} = P_{\NoExt{}}}$ (see the right-hand panel of \cref{fig:extDistHA-}).}
	Indeed, we see that for the vast majority of SFR tracers, and within the uncertainties of the measurements, ${\relDur{} \gtrsim 1}$ before this transition line and ${\relDur{} \lesssim 1}$ after, confirming our understanding of the behaviour seen in \cref{fig:scaleHA-,fig:scaleHA+,fig:scaleUV}.
	\cref{fig:scaleHA+,fig:scaleUV} show that the transition between both regimes shifts to higher levels of extinction for wider H$\alpha$ filters and redder UV wavelengths. For both of these, this shift is caused by an increasing contribution from the old stellar population, which has a flatter (i.e.\ more top-heavy) luminosity function than the H$\alpha$ luminosity function and therefore requires a larger amount of extinction to reach a systematic drop of the emergent flux.

	In relation to observational applications, the results presented here represent the maximum extent by which the SFR reference time-scale can be altered by extinction.
	With future simulations, that have feedback mechanisms to remove gas more effectively from the stellar birth environments, the values of \relDur{} that we measure will become more representative of real observations.
	However, different galaxies suffer from different levels of extinction, implying that deriving a single, uniform calibration of the reference time-scale under the influence of extinction remains challenging.
	If observed SFR tracer maps cannot be corrected for extinction, our results provide the extreme factor relative to unity by which the SFR tracer emission time-scale should be corrected.

	For our simulated galaxy, we find that \relDur{} and extinction do not have a simple monotonic relation.
	Extinction can both increase and decrease the duration of the SFR tracer emission time-scale.
	When comparing the results for the H${\alpha}$ filters \cref{fig:scaleHA-,fig:scaleHA+} and UV filters \cref{fig:scaleUV}, we can see that the extremes in \relDur{} are greater for UV than they are for H${\alpha}$.
    This means that UV filters are affected by extinction more strongly than H${\alpha}$.
	For real galaxies, which suffer from less extinction than permitted by the inefficient feedback model used in the simulation, not all star forming regions are still embedded in gas and \relDur{} must deviate from unity less than what we find here.

	\section{Sensitivity limit}\label{sec:sensitivity}
	Real-Universe observations are affected by sensitivity limits, which result from a combination of the telescope's intrinsic sensitivity and as part of the post-processing pipeline.
	The emission maps we have created so far recover all emission and do not have a sensitivity limit.
	Since extinction reduces the overall emission in an emission map, a sensitivity limit will more greatly affect the extincted emission maps compared to the unextincted emission maps.
	This will likely lead to lower SFR tracer emission time-scales (or values of \relDur{}).

	To investigate how \relDur{} changes with the sensitivity limit, we create new emission maps (both extincted and unextincted) by applying a flux density cut to our existing (${F_{\mathrm{d}} = 1}$) emission maps.
	The flux density cut is applied by setting to zero all the pixels in the emission map that are below the sensitivity limit.
	For each emission filter, the sensitivity limits are chosen as ${\left[10^{\left(E^{\max}_{\NoExt{}}  - 5\right)},~10^{\left(E^{\max}_{\NoExt{}}  - 2\right)}\right]}$ in 0.25 logarithmic steps, where ${E^{\max}_{\NoExt{}}}$ is the base-ten logarithm of the maximum emission in the \emph{unextincted} emission map.
	Choosing more extreme limits would be of little use.
	A larger upper limit would reduce the amount of emission within the extincted map to the extent that we could no longer use \textsc{Heisenberg} reliably, due to fewer than 35 regions being identified \citep[see][Section~4.4]{KRUI18}.
	A smaller lower limit would not add any new information, since \relDur{} will saturate at the values seen at ${F_{\mathrm{d}} = 1}$ in \cref{fig:scaleHA-,fig:scaleHA+,fig:scaleUV}.
	When running these experiments, we adopt the same sensitivity limits for both the extincted and unextincted emission maps.

	\begin{figure}
		\centering
		\includegraphics[width=\columnwidth]{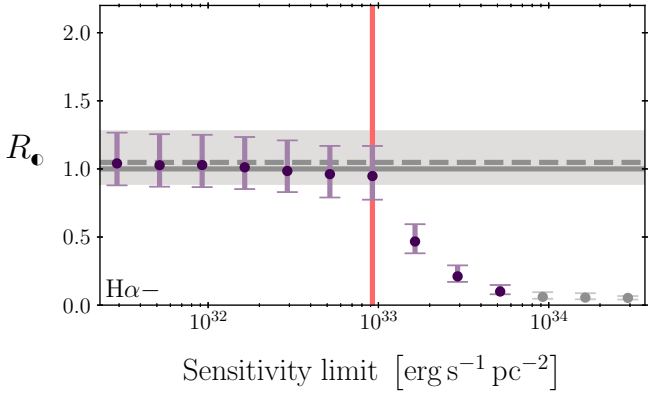}%
		\caption{
			Relative amount by which the continuum-subtracted H${\alpha}$ (H${\alpha{-}}$) emission time-scale changes due to extinction and a non-zero sensitivity limit.
			Shown is the relative duration of the time-scale associated to the extincted (${F_{\mathrm{d}} = 1}$, see \cref{sec:extinction}) compared to the unextincted SFR tracer emission map, \relDur{}, as a function of the sensitivity limit.
			The horizontal grey solid line (at ${\relDur{} = 1}$) indicates the point where the time-scale associated to each of the two emission maps is the same.
			The horizontal grey dashed line indicates the value of \relDur{} without a sensitivity limit applied (see \cref{fig:scaleHA-}), representing the expected saturation level.
			The grey-shaded region marks the uncertainty on this value.
			The vertical red line indicates the lowest contour level used by \textsc{Heisenberg} to identify emission peaks.
			The grey-shaded data points have fewer than 35 identified emission peaks in the extincted emission image and are therefore unreliable  \citep[see][Section~4.4]{KRUI18}.
		}\label{fig:limitHA-}
	\end{figure}
	\begin{figure}
		\centering
		\includegraphics[width=\columnwidth]{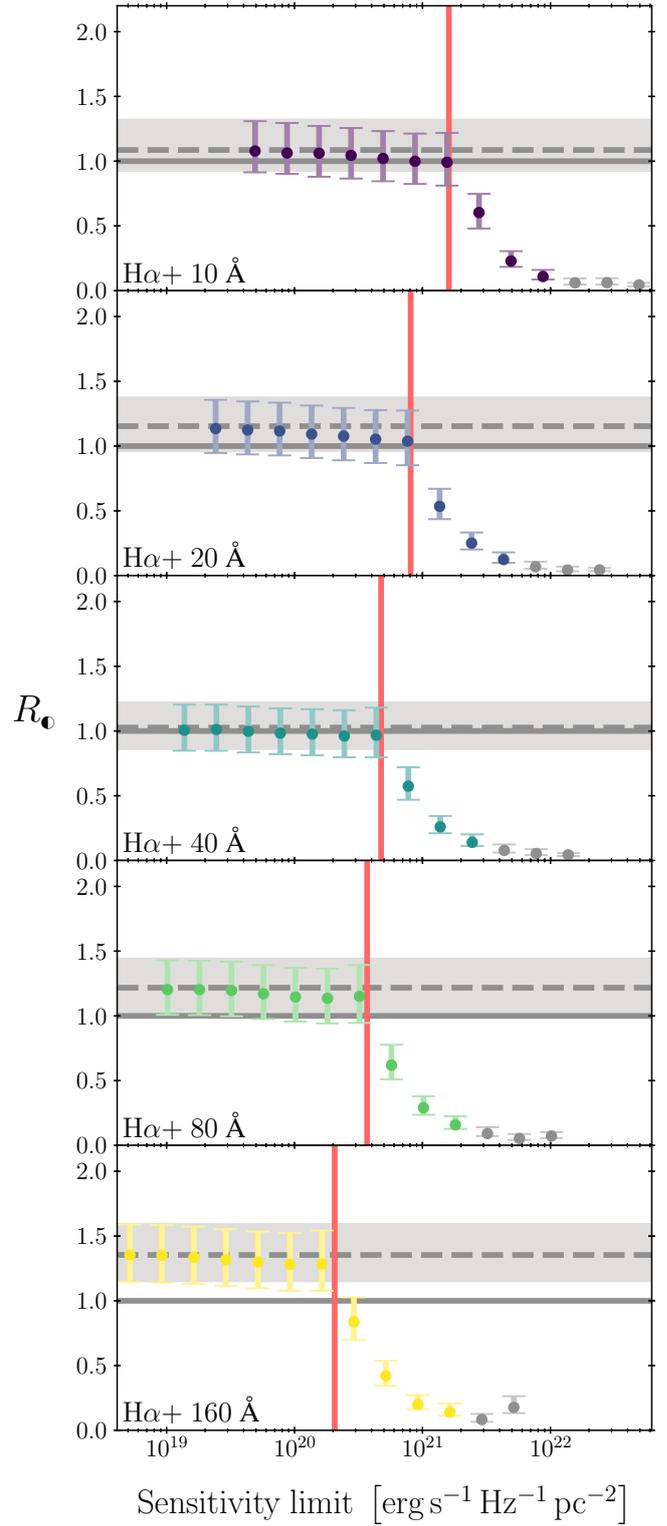}
		\caption{
			Same as \cref{fig:limitHA-} for H${\alpha}$ filters without continuum subtraction (H${\alpha{+}}$).
			The filter width is indicated in the bottom left corner of each panel.
		}\label{fig:limitHA+}
	\end{figure}
	\begin{figure*}
		\centering
		\includegraphics[width=0.9\textwidth]{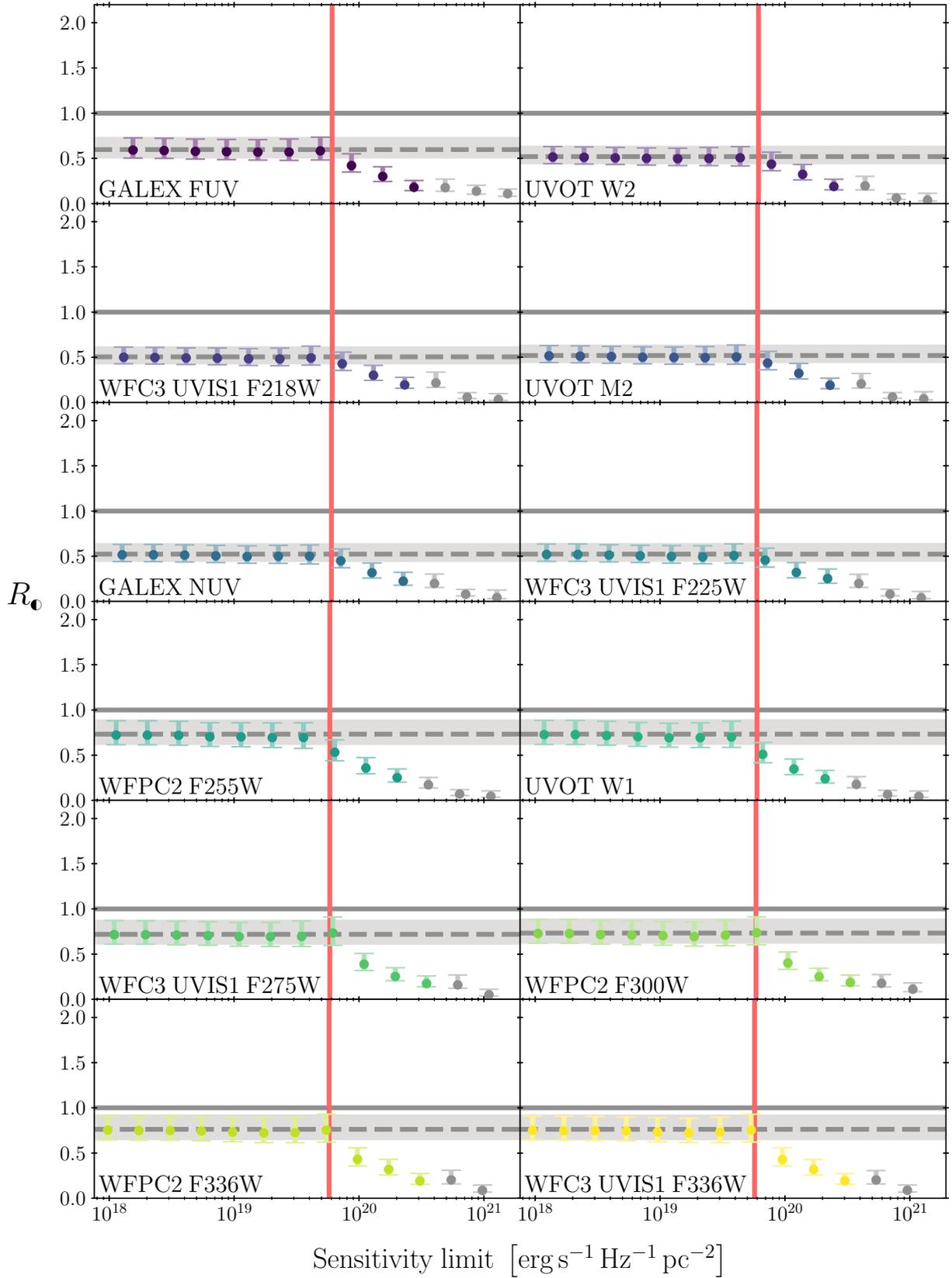}%
		\caption{
			Same as \cref{fig:limitHA-} for UV emission filters.
			The panels are ordered from left to right, top to bottom, by the filter-weighted mean wavelength.
			The filter name is indicated in the bottom left corner of each panel.
		}\label{fig:limitUV}
	\end{figure*}

	In \cref{fig:limitHA-,fig:limitHA+,fig:limitUV}, we show how \relDur{} changes with sensitivity limit, for the H${\alpha{-}}$, H${\alpha{+}}$, and UV filters.
	The figures show the behaviour that we expected: with a high sensitivity limit the value of \relDur{} is low (${\sim \text{0.1--0.2}}$) and increases (up to the saturation level) as the sensitivity limit decreases.
	The saturation level (the \relDur{} values seen for ${F_{\mathrm{d}} = 1}$ in \cref{sec:extinction}) is recovered for sensitivity limits up to the flux level down to which \textsc{Heisenberg} identifies emission peaks (about ${10^{33}~\mathrm{erg\,s^{-1}\,pc^{-2}}}$ for H${\alpha{-}}$ at the native column density of the simulation, i.e.\ ${F_{\mathrm{d}}=1}$).
	We also see that for sensitivity limits which are too high (${\gtrsim 10^{34}~\mathrm{erg\,s^{-1}\,pc^{-2}}}$ for H${\alpha{-}}$) remove too much emission and leave an insufficient number of peaks for the application of our analysis (grey data points).

	The reported values of \relDur{}, as a function of the sensitivity limit, are dependent on the results in the previous section, where we varied the column density scaling factor.
	If we had chosen to use a different density scaling factor, ${F_{\mathrm{d}}}$, these changes would also manifest themselves here.
	The most obvious way in which ${F_{\mathrm{d}}}$ affects the behaviour of \relDur{} as a function of the sensitivity limit is the value at which the measurements would saturate, because this simply matches the corresponding ${F_{\mathrm{d}}}$ measurement in \cref{sec:extinction}.
	The sensitivity limit at which the saturation level is reached is less obvious.
	In \cref{fig:limitHA-,fig:limitHA+,fig:limitUV}, we have included a line marking ${10^{\left(E^{\max}_{\Ext{}} - \Delta\right)}}$ where ${E^{\max}_{\Ext{}}}$ is the base-ten logarithm of the maximum emission in the \emph{extincted} emission map and ${\Delta = 2}$ (see \cref{sec:KL14} for details).
	This line is the lowest contour level used in our experiments by \textsc{Heisenberg} to identify emission peaks (across both input maps).
	The lowest contour level will become lower as ${F_{\mathrm{d}}}$ becomes higher (corresponding to more extinction).
	We expect, and observe, that \relDur{} saturates if the sensitivity limit is lower than the lowest contour level used by \textsc{Heisenberg}, allowing the same number of peaks to be identified.
	Above this level, emission is being removed that could be identified as emission peaks by \textsc{Heisenberg} and so reduces the apparent duration of the emission map.

	In summary, the results of this section show that if the sensitivity limit of the observations is lower than the lowest flux level that would be used by \textsc{Heisenberg} to identify emission peaks (typically chosen to be ${\Delta = 2}$~dex below the maximum; although, this is not always the case: for example values used in observations see \citealt{KRUI19,CHEV20}), \relDur{} will be unchanged relative to what would be measured without a sensitivity limit.
	Above this limit, the measured SFR tracer emission time-scale is reduced: the sensitivity limit further compounds the effects of extinction.
	
	\section{Application to observations} \label{sec:obs}
	
    \begin{table*}
      \caption{Maximum possible effect of extinction on the molecular cloud lifecycles inferred for 10 nearby galaxies from \citet{KRUI19} and \citet{CHEV20}. The first column lists the galaxy name, the second to fifth columns list the input quantities (total gas surface density, metallicity, implied gas surface density scaling factor, and log critical H$\alpha$ flux density threshold), and the sixth to eleventh columns list the results (minimum and maximum relative change due to extinction of the reference time-scale, molecular cloud lifetime, and feedback time-scale).} 
    \label{tab:apply}
      \begin{tabular*}{\textwidth}{l @{\extracolsep{\fill}} c c c c c c c c c c}
       \hline
       Galaxy & $\Sigma_{\rm g}$ & $12+\log({\rm O/H})$ & $F_{\rm d}$ & $\log_{10}\cal{F}_{\rm H\alpha,crit}$ & $\relDur{}_{\rm ,min}$ & $\relDur{}_{\rm ,max}$ & $t_{\rm CO,min}$ & $t_{\rm CO,max}$ & $t_{\rm fb,min}$ & $t_{\rm fb,max}$ \\ 
        & $[{\mathrm{M_{\odot}\,pc^{-2}}}]$ & [--] & [--] & $[{\rm erg}~{\rm s}^{-1}~{\rm pc}^{-2}]$ & [--] & [--] & [Myr] & [Myr] & [Myr] & [Myr] \\ 
       \hline
       NGC300 & $10.5$ & $8.36$ & $0.29$ & $31.69$ & $1.00$ & $1.16$ & $10.8^{+2.1}_{-1.7}$ & $12.6^{+2.4}_{-2.0}$ & $1.5^{+0.2}_{-0.2}$ & $1.7^{+0.2}_{-0.2}$ \\ 
       NGC628 & $20.9$ & $8.66$ & $1.17$ & $32.53$ & $0.87$ & $1.00$ & $20.8^{+3.1}_{-2.2}$ & $24.0^{+3.6}_{-2.5}$ & $2.8^{+0.5}_{-0.3}$ & $3.2^{+0.6}_{-0.4}$ \\ 
       NGC3351 & $7.3$ & $8.75$ & $0.50$ & $32.41$ & $1.00$ & $1.18$ & $20.6^{+3.4}_{-3.0}$ & $24.3^{+4.0}_{-3.5}$ & $2.5^{+0.8}_{-0.6}$ & $2.9^{+0.9}_{-0.7}$ \\ 
       NGC3627 & $15.6$ & $8.63$ & $0.82$ & $32.79$ & $1.00$ & $1.10$ & $18.9^{+3.4}_{-3.2}$ & $20.8^{+3.7}_{-3.5}$ & $2.8^{+0.8}_{-0.7}$ & $3.1^{+0.9}_{-0.8}$ \\ 
       NGC4254 & $38.0$ & $8.68$ & $2.22$ & $32.95$ & $0.37$ & $1.00$ & $7.7^{+1.4}_{-0.8}$ & $20.9^{+3.9}_{-2.3}$ & $1.8^{+0.4}_{-0.4}$ & $4.8^{+1.1}_{-1.0}$ \\ 
       NGC4303 & $35.8$ & $8.63$ & $1.88$ & $32.52$ & $0.43$ & $1.00$ & $7.2^{+2.0}_{-0.9}$ & $16.9^{+4.6}_{-2.2}$ & $1.7^{+0.8}_{-0.4}$ & $4.0^{+1.8}_{-1.0}$ \\ 
       NGC4321 & $13.5$ & $8.70$ & $0.83$ & $32.57$ & $1.00$ & $1.09$ & $19.1^{+2.3}_{-2.2}$ & $20.9^{+2.5}_{-2.4}$ & $3.3^{+0.7}_{-0.6}$ & $3.6^{+0.8}_{-0.7}$ \\ 
       NGC4535 & $16.8$ & $8.62$ & $0.86$ & $32.89$ & $1.00$ & $1.09$ & $26.4^{+4.7}_{-3.6}$ & $28.7^{+5.1}_{-3.9}$ & $3.9^{+1.2}_{-0.9}$ & $4.2^{+1.3}_{-1.0}$ \\ 
       NGC5068 & $11.1$ & $8.46$ & $0.39$ & $32.70$ & $1.00$ & $1.17$ & $9.6^{+2.9}_{-1.8}$ & $11.2^{+3.4}_{-2.1}$ & $1.0^{+0.4}_{-0.3}$ & $1.2^{+0.5}_{-0.4}$ \\ 
       NGC5194 & $69.6$ & $8.84$ & $5.88$ & $31.64$ & $0.14$ & $1.00$ & $4.4^{+1.3}_{-0.7}$ & $30.5^{+9.2}_{-4.8}$ & $0.7^{+0.3}_{-0.2}$ & $4.8^{+2.1}_{-1.1}$ \\ 
       \hline
      \end{tabular*} 
    \end{table*}

	We now apply the results of this paper to the measurements of the molecular cloud lifecycle in ten nearby galaxies from \citet{KRUI19} and \citet{CHEV20}. The corresponding H$\alpha$ maps have been taken with the WFI instrument on the MPG/ESO 2.2-m telescope at La Silla Observatory (NGC300, NGC628, NGC3627, NGC4254, NGC4303, NGC4535, and NGC5068; \citealt{FAES14}, Razza et al.~in prep.) and the Kitt Peak National Observatory 2.1m telescope with the CFIM imager (NGC3351, NGC4321, NGC5194; \citealt{KENN03}). The maps are continuum-subtracted using $R$-band imaging. They also include a uniform correction for contamination by [N{\sc ii}] line emission, based on integral field spectroscopic observations of the same targets, assuming a contamination of 20~per~cent for NGC300 \citep{McLE20} and 30~per~cent for the other galaxies \citep{KREC19}. The maps are not corrected for stellar absorption, but its effect is negligible for the young regions that we consider \citep{HAYD18}.
	
	The results of this exercise are shown in \cref{tab:apply}. The first set of columns show the input quantities, i.e.\ the kpc-scale total gas surface density, the metallicity, and the resulting scaling factor $F_{\mathrm{d}}$, which we define as
	\begin{equation}
	    \label{eq:fd}
	    F_{\mathrm{d}} = \frac{\Sigma_{\rm g}}{\Sigma_{\rm g,0}}\times10^{\log({\rm O/H})-\log({\rm O/H})_\odot}~,
	\end{equation}
	with $\Sigma_{\rm g,0}=16.7~{\mathrm{M_{\odot}\,pc^{-2}}}$ (see \cref{fig:ColDen}) and $12+\log({\rm O/H})_\odot=8.69$ \citep{ASPL09}. This definition normalises the observed total gas surface density to the surface density in the simulation, and applies an additional metallicity scaling to reflect the linear increase of the dust mass with the metal fraction. We also list the H$\alpha$ flux density threshold used to mask noise in the adopted H$\alpha$ maps ($\cal{F}_{\rm H\alpha,crit}$). This threshold always falls below the critical value above which \relDur{} deviates from unity (see \cref{fig:limitHA-}), which means that the adopted H$\alpha$ maps are sufficiently sensitive for our analysis. Using \cref{fig:scaleHA-}, we then estimate the upper and lower limits on \relDur, as well as the implied upper and lower limits on the molecular cloud lifetime ($t_{\rm CO}$) and on the feedback time-scale ($t_{\rm fb}$), i.e.\ the time for which CO and H$\alpha$ emission coexist and represents the time taken by massive stars to disperse their host molecular cloud.
	
	\cref{tab:apply} shows that most of the galaxies show a small change in the reference time-scale due to extinction, resulting in changes to the molecular cloud lifecycle by roughly the quoted uncertainties (or less) in seven out of ten galaxies. In six out of ten cases, the gas surface density is low ($\Sigma_{\rm g}\la20~{\mathrm{M_{\odot}\,pc^{-2}}}$) and extinction may compress the dynamic range of the region luminosities, so that the adopted reference time-scale is a minimum -- these are the rows with $\relDur{}_{\rm ,min}=1.00$, implying that $t_{\rm CO,min}$ and $t_{\rm fb,min}$ show the original time-scales that do not account for extinction, whereas $t_{\rm CO,max}$ and $t_{\rm fb,max}$ show how they may change under the maximum effect of extinction. The maximum effect is of the order 20~per~cent, which falls within the typical uncertainties. In four out of ten cases, the gas surface density is high ($\Sigma_{\rm g}\ga20~{\mathrm{M_{\odot}\,pc^{-2}}}$) and extinction may shorten the reference time-scale -- these are the rows with $\relDur{}_{\rm ,max}=1.00$, implying that $t_{\rm CO,max}$ and $t_{\rm fb,max}$ show the original time-scales that do not account for extinction, whereas $t_{\rm CO,min}$ and $t_{\rm fb,min}$ show how they may change under the maximum effect of extinction. The dynamic range of the relative change in these cases is larger than at low gas surface densities, spanning $\relDur{}_{\rm ,min}=0.14{-}0.87$.
	
	As a result of extinction, the molecular cloud lifecycle in NGC4254 and NGC4303 may proceed faster by up to a factor of 2.3--2.7 than reported by \citet{CHEV20}. However, fig.~5 of that paper shows that the observed cloud lifetimes are largely consistent with their internal dynamical time-scales, implying that a shorter time-scale could be unphysical. It is plausible that the true impact of extinction is smaller, because the relative changes reported in \cref{tab:apply} represent the maximum possible effect of extinction, due to the fact that the feedback model in the used simulation is too ineffective at dispersing molecular clouds. However, it is possible that the long cloud lifetime in the central region of NGC4254 relative to the cloud dynamical time (see the left-most data point in the middle-right panel of fig.~5 in \citealt{CHEV20}) results from extinction -- decreasing this cloud lifetime by a factor of 2 would make it consistent with a dynamical time. None the less, it is unlikely that extinction significantly affects the galaxy-wide results for these two galaxies.
	
	Likewise, the molecular cloud lifetimes in NGC5194 (M51) reported by \citet{CHEV20} fall about a factor of 2.5 (or 0.4~dex) above the time-scales expected due to galactic dynamics \citep[including midplane free-fall, epicyclic perturbations, shear, cloud-cloud collisions, and spiral arm passages, using the model of][]{JEFF18}. We now see that including the effects of extinction may decrease the cloud lifetimes by up to a factor of 7 (or 0.8~dex). While the maximum effect of extinction is overestimated here, it is clear that the molecular cloud lifetime in NGC5194 is potentially consistent with the expected time-scale due to galactic dynamical processes.
	
	In summary, we find that the effects of extinction can lead to quantitative changes in some galaxies, but likely do not qualitatively change the measurements made previously in most systems. The molecular cloud lifetimes remain short, of the order a dynamical time, even after applying the effects of extinction. A similar conclusion extends to the feedback time-scales. The measurements of \citet{KRUI19} and \citet{CHEV20} imply that early, pre-supernova feedback leads to the dispersal of molecular clouds. The effects of extinction only strengthen this conclusion because galaxies that are potentially the most strongly affected by extinction due to their high gas surface densities generally have the longest feedback time-scales. Including the effects of extinction may shorten the dispersal time-scales convincingly into the regime where supernovae have not detonated yet. However, as before, this represents the extreme case. It is possible that the affected feedback time-scales show considerably less change.

	\section{Conclusions}\label{sec:conclusion}
	Using a new statistical method (the \enquote{uncertainty principle for star formation}, used through the \textsc{Heisenberg} code \citealt{KRUI14,KRUI18}), we determine how the apparent characteristic emission time-scale of SFR tracers changes as a result of dust extinction.
	In previous work \citep{HAYD18} we characterised how the durations over which H${\alpha}$ and UV emission emerges from coeval stellar populations changes as a function of the metallicity and the SFR surface density (to emulate incomplete IMF sampling); however, we excluded the effects of extinction.
	In the present work, we include the effects of extinction.
	Extinction does not change the underlying characteristic time-scale of the emission, but we demonstrate that it changes the flux density distribution of the observed emission.
	As a result, the apparent duration of the emission time-scale measured with the \textsc{Heisenberg} code is changed and therefore the reference time-scale required for calibrating the evolutionary time-line of observational applications of the \textsc{Heisenberg} code.
	We quantify this change through \relDur{}, the relative duration of the time-scale associated to a SFR tracer emission map affected by extinction relative to the time-scale associated to the SFR tracer emission map without extinction.

	For the analysis in this paper, we used a Milky-Way-like disc galaxy simulation \citep{FUJI18,FUJI19} from which we generate synthetic SFR tracer emission maps.
	This \textsc{enzo} simulation has the advantage that on large scales (${\gtrsim 100}$~pc) it agrees with observational constraints (e.g.\ the global- and kpc-scale Kennicutt-Schmidt relations and the phases of the interstellar medium); however, as we demonstrate in \citet{FUJI19}, the implemented feedback mechanisms are insufficient to disperse molecular clouds surrounding sites of star formation, in clear contradiction with observations \citep[e.g.][]{KRUI19,CHEV20}.
	This means that this simulation acts as a limiting case of the effects of extinction: the magnitude of the effects we find in this paper represent the most extreme case.
	In real-Universe observational studies, the effects of extinction must be less.
	We are therefore able to use the results here to indicate when extinction could have an impact on our measurements for real observations and the maximum extent that they could be affected.
	For conditions where we find that extinction does not affect the SFR tracer reference time-scale, observational applications are firmly ruled out to be affected by extinction.

	We produce synthetic emission maps of 18 SFR tracer filters: 12 different UV filters, 5 H${\alpha}$ filters with continuum (H${\alpha{+}}$) and a continuum subtracted H${\alpha}$ filter (H${\alpha{-}}$).
	To generate our synthetic emission maps, we use the properties of the stars within the simulation along with the stochastic stellar population synthesis code \textsc{slug2} \citep{SILV12, SILV14, KRUM15} to determine the emission spectrum associated to each of the stars present.
	The amount of extinction affecting each star comes from the properties of the gas present within the simulation that falls within the line of sight of the star.
	This information, along with a Milky Way extinction curve, is used by \textsc{slug2} to produce the extincted synthetic emission maps.
	In this work we only consider solar metallicity and a fully sampled IMF (see below).

	Our experiments show that extinction affects \relDur{} non-monotonically.
	At low levels of extinction, corresponding to mean galactic surface densities ${\Sigma_{\mathrm{g}} \lesssim 20~\mathrm{M_{\odot}\,pc^{-2}}}$, \relDur{} is close to but slightly larger than unity, because only light from the brightest star-forming regions, which are in the regions of highest extinction, suffers any significant attenuation.
	This compresses the range of fluxes present in the extincted map relative to the unextincted one, increasing the total number of distinct regions within a fixed dynamic range, an effect that \textsc{Heisenberg}'s statistical analysis interprets as a slight increase in the time-scale of the extincted map compared to the unextincted one.
	As the extinction increases and most star-forming regions begin to suffer significant attenuation, the effect reverses and \relDur{} becomes smaller than unity by as much as a factor of ten in the most extreme case.

	We also investigate how a sensitivity limit on our observations could impact \relDur{}.
	The critical value at which the sensitivity limit has an impact on \relDur{} depends on the lowest flux level used by \textsc{Heisenberg} to identify emission peaks, which is typically chosen to be 2~dex below the maximum flux level in the emission map \citep[see e.g.][]{KRUI18,KRUI19,CHEV20}.
	If the sensitivity limit is lower than this threshold, \relDur{} is unchanged from the results we recover if there was no sensitivity limit.
	For sensitivity limits above the threshold, \relDur{} can decrease to as low as ${\sim \text{0.1--0.2}}$.
	
	In addition, we apply the insights drawn from this paper to the observational characterisation of the molecular cloud lifecycle by \citet{KRUI19} and \citet{CHEV20}. Most galaxies have sufficiently low gas surface densities that the maximal effect of extinction leads to only quantitative changes that are similar to or fall within the uncertainties on the cloud lifetimes and feedback time-scales. The three galaxies with the highest kpc-scale gas surface densities ($\Sigma_{\rm g}\ga20~{\mathrm{M_{\odot}\,pc^{-2}}}$; NGC4254, NGC4303, and NGC5194) could potentially have overestimated molecular cloud lifetimes and feedback time-scales. Broadly speaking, correcting for the effects of extinction goes in the direction of making the measured time-scales more consistent with the general conclusions drawn by \citet{KRUI19} and \citet{CHEV20}, i.e.\ that molecular clouds live for a dynamical time (which can either be the galactic or internal dynamical time-scale, depending on the kpc-scale gas surface density, see \citealt{CHEV20}) and are dispersed by early, pre-supernova feedback.

	In summary, we have measured the relative change in the characteristic emission time-scales for SFR tracers (H${\alpha}$ and UV) as a result of extinction and how this might be compounded by the non-zero sensitivity limits of observational data.
	We find that extinction does not affect the SFR tracer emission time-scales at gas surface densities ${\Sigma_{\mathrm{g}} \lesssim 20~\mathrm{M_{\odot}\,pc^{-2}}}$.
	At higher surface densities, the SFR tracer emission time-scales may be reduced (relative to the values presented in \citealt{HAYD18}) by a factor as low as ${\relDur{} = 0.1}$.
	However, the values presented in this work represent extreme limits.
	The ineffectiveness of stellar feedback at dispersing molecular clouds in the simulation used here has enabled us to set limits on the impact of feedback on SFR tracer time-scales.
	The resulting correction factors are critical for informing observational efforts characterising the molecular cloud lifecycle using the \textsc{Heisenberg} code.
	Future calibrations using simulations with improved feedback physics will provide a better description of the embedded phase of massive star formation and will thus be able to further improve the limits provided here.

	\section*{Acknowledgements}
	We thank an anonymous referee for a constructive and insightful report, which improved this paper.
	DTH is a fellow of the International Max Planck Research School for Astronomy and Cosmic Physics at the University of Heidelberg (IMPRS-HD).
	DTH, YF, MC, JMDK, and MRK acknowledge support from the Australia-Germany Joint Research Cooperation Scheme (UA-DAAD, grant number 57387355).
	MC and JMDK gratefully acknowledge funding from the German Research Foundation (DFG) in the form of an Emmy Noether Research Group (grant number KR4801/1-1) and the DFG Sachbeihilfe (grant number KR4801/2-1).
	JMDK gratefully acknowledges funding from the European Research Council (ERC) under the European Union's Horizon 2020 research and innovation programme via the ERC Starting Grant MUSTANG (grant agreement number 714907).
	MRK acknowledges support from the Australia Research Council's Discovery Projects and Future Fellowship funding schemes, awards DP160100695 and FT180100375. MRK acknowledges support from the Alexander von Humboldt Foundation in the form of a Humboldt Research Award.
	This research was undertaken with the assistance of resources from the National Computational Infrastructure (NCI Australia), an NCRIS enabled capability supported by the Australian Government.

    \section*{Data availability}
    The data underlying this article will be shared on reasonable request to the corresponding author.
	%%%%%%%%%%%%%%%%%%%%%%%%%%%%%%%%%%%%%%%%%%%%%%%%%%



	\bibliographystyle{mnras}
	\bibliography{bib/bibliography.bib} % if your bibtex file is called example.bib

\begin{thebibliography}{}
\makeatletter
\relax
\def\mn@urlcharsother{\let\do\@makeother \do\$\do\&\do\#\do\^\do\_\do\%\do\~}
\def\mn@doi{\begingroup\mn@urlcharsother \@ifnextchar [ {\mn@doi@}
  {\mn@doi@[]}}
\def\mn@doi@[#1]#2{\def\@tempa{#1}\ifx\@tempa\@empty \href
  {http://dx.doi.org/#2} {doi:#2}\else \href {http://dx.doi.org/#2} {#1}\fi
  \endgroup}
\def\mn@eprint#1#2{\mn@eprint@#1:#2::\@nil}
\def\mn@eprint@arXiv#1{\href {http://arxiv.org/abs/#1} {{\tt arXiv:#1}}}
\def\mn@eprint@dblp#1{\href {http://dblp.uni-trier.de/rec/bibtex/#1.xml}
  {dblp:#1}}
\def\mn@eprint@#1:#2:#3:#4\@nil{\def\@tempa {#1}\def\@tempb {#2}\def\@tempc
  {#3}\ifx \@tempc \@empty \let \@tempc \@tempb \let \@tempb \@tempa \fi \ifx
  \@tempb \@empty \def\@tempb {arXiv}\fi \@ifundefined
  {mn@eprint@\@tempb}{\@tempb:\@tempc}{\expandafter \expandafter \csname
  mn@eprint@\@tempb\endcsname \expandafter{\@tempc}}}

\bibitem[\protect\citeauthoryear{{Asplund}, {Grevesse}, {Sauval}  \&
  {Scott}}{{Asplund} et~al.}{2009}]{ASPL09}
{Asplund} M.,  {Grevesse} N.,  {Sauval} A.~J.,   {Scott} P.,  2009, \mn@doi
  [\araa] {10.1146/annurev.astro.46.060407.145222}, \href
  {http://adsabs.harvard.edu/abs/2009ARA%26A..47..481A} {47, 481}

\bibitem[\protect\citeauthoryear{Berman}{Berman}{1936}]{BERM36}
Berman L.,  1936, \mn@doi [\mnras] {10.1093/mnras/96.9.890}, \href
  {https://ui.adsabs.harvard.edu/abs/1936MNRAS..96..890B} {96, 890}

\bibitem[\protect\citeauthoryear{Bryan et~al.,}{Bryan et~al.}{2014}]{BRYA14}
Bryan G.~L.,  et~al., 2014, \mn@doi [\apjs] {10.1088/0067-0049/211/2/19}, \href
  {https://ui.adsabs.harvard.edu/abs/2014ApJS..211...19B} {211, 19}

\bibitem[\protect\citeauthoryear{Calzetti, Kinney  \&
  Storchi-Bergmann}{Calzetti et~al.}{1994}]{CALZ94}
Calzetti D.,  Kinney A.~L.,   Storchi-Bergmann T.,  1994, \mn@doi [\apj]
  {10.1086/174346}, \href
  {https://ui.adsabs.harvard.edu/abs/1994ApJ...429..582C} {429, 582}

\bibitem[\protect\citeauthoryear{Chabrier}{Chabrier}{2005}]{CHAB05}
Chabrier G.,  2005, in Corbelli E.,  Palla F.,   Zinnecker H.,  eds,
  Astrophysics and Space Science Library Vol. 327, The Initial Mass Function 50
  Years Later. p.~41 (\mn@eprint {arXiv} {astro-ph/0409465}),
  \mn@doi{10.1007/978-1-4020-3407-7_5}

\bibitem[\protect\citeauthoryear{{Chevance} et~al.,}{{Chevance}
  et~al.}{2020}]{CHEV20}
{Chevance} M.,  et~al., 2020, \mn@doi [\mnras] {10.1093/mnras/stz3525}, \href
  {https://ui.adsabs.harvard.edu/abs/2020MNRAS.493.2872C} {493, 2872}

\bibitem[\protect\citeauthoryear{Draine \& Bertoldi}{Draine \&
  Bertoldi}{1996}]{DRAI96}
Draine B.~T.,  Bertoldi F.,  1996, \mn@doi [\apj] {10.1086/177689}, \href
  {https://ui.adsabs.harvard.edu/#abs/1996ApJ...468..269D} {468, 269}

\bibitem[\protect\citeauthoryear{{Faesi}, {Lada}, {Forbrich}, {Menten}  \&
  {Bouy}}{{Faesi} et~al.}{2014}]{FAES14}
{Faesi} C.~M.,  {Lada} C.~J.,  {Forbrich} J.,  {Menten} K.~M.,   {Bouy} H.,
  2014, \mn@doi [\apj] {10.1088/0004-637X/789/1/81}, \href
  {https://ui.adsabs.harvard.edu/abs/2014ApJ...789...81F} {789, 81}

\bibitem[\protect\citeauthoryear{Fujimoto, Krumholz  \& Tachibana}{Fujimoto
  et~al.}{2018}]{FUJI18}
Fujimoto Y.,  Krumholz M.~R.,   Tachibana S.,  2018, \mn@doi [\mnras]
  {10.1093/mnras/sty2132}, \href
  {https://ui.adsabs.harvard.edu/abs/2018MNRAS.480.4025F} {480, 4025}

\bibitem[\protect\citeauthoryear{Fujimoto, Chevance, Haydon, Krumholz  \&
  Kruijssen}{Fujimoto et~al.}{2019}]{FUJI19}
Fujimoto Y.,  Chevance M.,  Haydon D.~T.,  Krumholz M.~R.,   Kruijssen J.
  M.~D.,  2019, \mn@doi [\mnras] {10.1093/mnras/stz641}, \href
  {https://ui.adsabs.harvard.edu/abs/2019MNRAS.487.1717F} {487, 1717}

\bibitem[\protect\citeauthoryear{Girardi, Bressan, Bertelli  \& Chiosi}{Girardi
  et~al.}{2000}]{GIRA00}
Girardi L.,  Bressan A.,  Bertelli G.,   Chiosi C.,  2000, \mn@doi [\aaps]
  {10.1051/aas:2000126}, \href
  {https://ui.adsabs.harvard.edu/abs/2000A&AS..141..371G} {141, 371}

\bibitem[\protect\citeauthoryear{Hao, Kennicutt, Johnson, Calzetti, Dale  \&
  Moustakas}{Hao et~al.}{2011}]{HAO11}
Hao C.-N.,  Kennicutt R.~C.,  Johnson B.~D.,  Calzetti D.,  Dale D.~A.,
  Moustakas J.,  2011, \mn@doi [\apj] {10.1088/0004-637X/741/2/124}, \href
  {https://ui.adsabs.harvard.edu/abs/2011ApJ...741..124H} {741, 124}

\bibitem[\protect\citeauthoryear{Haydon, Kruijssen, Hygate, Schruba, Krumholz,
  Chevance  \& Longmore}{Haydon et~al.}{2018}]{HAYD18}
Haydon D.~T.,  Kruijssen J. M.~D.,  Hygate A. P.~S.,  Schruba A.,  Krumholz
  M.~R.,  Chevance M.,   Longmore S.~N.,  2018, preprint, \href
  {https://ui.adsabs.harvard.edu/#abs/2018arXiv181010897H} {p.
  arXiv:1810.10897} (\mn@eprint {arXiv} {1810.10897})

\bibitem[\protect\citeauthoryear{James, Shane, Knapen, Etherton  \&
  Percival}{James et~al.}{2005}]{JAME05}
James P.~A.,  Shane N.~S.,  Knapen J.~H.,  Etherton J.,   Percival S.~M.,
  2005, \mn@doi [\aap] {10.1051/0004-6361:20035892}, \href
  {https://ui.adsabs.harvard.edu/abs/2005A&A...429..851J} {429, 851}

\bibitem[\protect\citeauthoryear{{Jeffreson} \& {Kruijssen}}{{Jeffreson} \&
  {Kruijssen}}{2018}]{JEFF18}
{Jeffreson} S.~M.~R.,  {Kruijssen} J.~M.~D.,  2018, \mn@doi [\mnras]
  {10.1093/mnras/sty594}, \href
  {http://adsabs.harvard.edu/abs/2018MNRAS.476.3688J} {476, 3688}

\bibitem[\protect\citeauthoryear{Kawamura et~al.,}{Kawamura
  et~al.}{2009}]{KAWA09}
Kawamura A.,  et~al., 2009, \mn@doi [\apjs] {10.1088/0067-0049/184/1/1}, \href
  {https://ui.adsabs.harvard.edu/abs/2009ApJS..184....1K} {184, 1}

\bibitem[\protect\citeauthoryear{Kennicutt \& Evans}{Kennicutt \&
  Evans}{2012}]{KENN12}
Kennicutt R.~C.,  Evans N.~J.,  2012, \mn@doi [\araa]
  {10.1146/annurev-astro-081811-125610}, \href
  {https://ui.adsabs.harvard.edu/abs/2012ARA&A..50..531K} {50, 531}

\bibitem[\protect\citeauthoryear{{Kennicutt} Robert~C. et~al.,}{{Kennicutt}
  et~al.}{2003}]{KENN03}
{Kennicutt} Robert~C. J.,  et~al., 2003, \mn@doi [\pasp] {10.1086/376941},
  \href {https://ui.adsabs.harvard.edu/abs/2003PASP..115..928K} {115, 928}

\bibitem[\protect\citeauthoryear{{Kreckel}, {Blanc}, {Schinnerer}, {Groves},
  {Adamo}, {Hughes}  \& {Meidt}}{{Kreckel} et~al.}{2016}]{KREC16}
{Kreckel} K.,  {Blanc} G.~A.,  {Schinnerer} E.,  {Groves} B.,  {Adamo} A.,
  {Hughes} A.,   {Meidt} S.,  2016, \mn@doi [\apj]
  {10.3847/0004-637X/827/2/103}, \href
  {https://ui.adsabs.harvard.edu/abs/2016ApJ...827..103K} {827, 103}

\bibitem[\protect\citeauthoryear{{Kreckel} et~al.,}{{Kreckel}
  et~al.}{2019}]{KREC19}
{Kreckel} K.,  et~al., 2019, \mn@doi [\apj] {10.3847/1538-4357/ab5115}, \href
  {https://ui.adsabs.harvard.edu/abs/2019ApJ...887...80K} {887, 80}

\bibitem[\protect\citeauthoryear{Kruijssen \& Longmore}{Kruijssen \&
  Longmore}{2014}]{KRUI14}
Kruijssen J. M.~D.,  Longmore S.~N.,  2014, \mn@doi [\mnras]
  {10.1093/mnras/stu098}, \href
  {https://ui.adsabs.harvard.edu/abs/2014MNRAS.439.3239K} {439, 3239}

\bibitem[\protect\citeauthoryear{{Kruijssen}, {Schruba}, {Hygate}, {Hu},
  {Haydon}  \& {Longmore}}{{Kruijssen} et~al.}{2018}]{KRUI18}
{Kruijssen} J.~M.~D.,  {Schruba} A.,  {Hygate} A.~P.~S.,  {Hu} C.-Y.,  {Haydon}
  D.~T.,   {Longmore} S.~N.,  2018, \mn@doi [\mnras] {10.1093/mnras/sty1128},
  \href {http://adsabs.harvard.edu/abs/2018MNRAS.479.1866K} {479, 1866}

\bibitem[\protect\citeauthoryear{{Kruijssen} et~al.,}{{Kruijssen}
  et~al.}{2019}]{KRUI19}
{Kruijssen} J.~M.~D.,  et~al., 2019, \mn@doi [\nat]
  {10.1038/s41586-019-1194-3}, \href
  {https://ui.adsabs.harvard.edu/abs/2019Natur.569..519K} {569, 519}

\bibitem[\protect\citeauthoryear{Krumholz, Fumagalli, da Silva, Rendahl  \&
  Parra}{Krumholz et~al.}{2015}]{KRUM15}
Krumholz M.~R.,  Fumagalli M.,  da Silva R.~L.,  Rendahl T.,   Parra J.,  2015,
  \mn@doi [\mnras] {10.1093/mnras/stv1374}, \href
  {https://ui.adsabs.harvard.edu/abs/2015MNRAS.452.1447K} {452, 1447}

\bibitem[\protect\citeauthoryear{Leitherer et~al.,}{Leitherer
  et~al.}{1999}]{LEIT99}
Leitherer C.,  et~al., 1999, \mn@doi [\apjs] {10.1086/313233}, \href
  {https://ui.adsabs.harvard.edu/abs/1999ApJS..123....3L} {123, 3}

\bibitem[\protect\citeauthoryear{McKee \& Williams}{McKee \&
  Williams}{1997}]{MCKE97}
McKee C.~F.,  Williams J.~P.,  1997, \mn@doi [\apj] {10.1086/303587}, \href
  {https://ui.adsabs.harvard.edu/abs/1997ApJ...476..144M} {476, 144}

\bibitem[\protect\citeauthoryear{{McLeod} et~al.,}{{McLeod}
  et~al.}{2020}]{McLE20}
{McLeod} A.~F.,  et~al., 2020, \mn@doi [\apj] {10.3847/1538-4357/ab6d63}, \href
  {https://ui.adsabs.harvard.edu/abs/2020ApJ...891...25M} {891, 25}

\bibitem[\protect\citeauthoryear{Safranek-Shrader, Krumholz, Kim, Ostriker,
  Klein, Li, McKee  \& Stone}{Safranek-Shrader et~al.}{2017}]{SAFR17}
Safranek-Shrader C.,  Krumholz M.~R.,  Kim C.-G.,  Ostriker E.~C.,  Klein
  R.~I.,  Li S.,  McKee C.~F.,   Stone J.~M.,  2017, \mn@doi [\mnras]
  {10.1093/mnras/stw2647}, \href
  {https://ui.adsabs.harvard.edu/abs/2017MNRAS.465..885S} {465, 885}

\bibitem[\protect\citeauthoryear{Schaller, Schaerer, Meynet  \&
  Maeder}{Schaller et~al.}{1992}]{SCHA92}
Schaller G.,  Schaerer D.,  Meynet G.,   Maeder A.,  1992, \aaps, \href
  {https://ui.adsabs.harvard.edu/abs/1992A&AS...96..269S} {96, 269}

\bibitem[\protect\citeauthoryear{Tamburro, Rix, Walter, Brinks, de Blok,
  Kennicutt  \& Mac~Low}{Tamburro et~al.}{2008}]{TAMB08}
Tamburro D.,  Rix H.~W.,  Walter F.,  Brinks E.,  de Blok W. J.~G.,  Kennicutt
  R.~C.,   Mac~Low M.~M.,  2008, \mn@doi [\aj] {10.1088/0004-6256/136/6/2872},
  \href {https://ui.adsabs.harvard.edu/abs/2008AJ....136.2872T} {136, 2872}

\bibitem[\protect\citeauthoryear{Tasker \& Tan}{Tasker \& Tan}{2009}]{TASK09}
Tasker E.~J.,  Tan J.~C.,  2009, \mn@doi [\apj] {10.1088/0004-637X/700/1/358},
  \href {https://ui.adsabs.harvard.edu/abs/2009ApJ...700..358T} {700, 358}

\bibitem[\protect\citeauthoryear{Toomre}{Toomre}{1964}]{TOOM64}
Toomre A.,  1964, \mn@doi [\apj] {10.1086/147861}, \href
  {https://ui.adsabs.harvard.edu/abs/1964ApJ...139.1217T} {139, 1217}

\bibitem[\protect\citeauthoryear{Turk, Smith, Oishi, Skory, Skillman, Abel  \&
  Norman}{Turk et~al.}{2011}]{TURK11}
Turk M.~J.,  Smith B.~D.,  Oishi J.~S.,  Skory S.,  Skillman S.~W.,  Abel T.,
  Norman M.~L.,  2011, \mn@doi [\apjs] {10.1088/0067-0049/192/1/9}, \href
  {https://ui.adsabs.harvard.edu/abs/2011ApJS..192....9T} {192, 9}

\bibitem[\protect\citeauthoryear{V{\'a}zquez \& Leitherer}{V{\'a}zquez \&
  Leitherer}{2005}]{VAZQ05}
V{\'a}zquez G.~A.,  Leitherer C.,  2005, \mn@doi [\apj] {10.1086/427866}, \href
  {https://ui.adsabs.harvard.edu/abs/2005ApJ...621..695V} {621, 695}

\bibitem[\protect\citeauthoryear{{Ward}, {Chevance}, {Kruijssen}, {Hygate},
  {Schruba}  \& {Longmore}}{{Ward} et~al.}{2020}]{WARD20}
{Ward} J.~L.,  {Chevance} M.,  {Kruijssen} J.~M.~D.,  {Hygate} A. P.~S.,
  {Schruba} A.,   {Longmore} S.~N.,  2020, \mn@doi [\mnras\ in press,
  arXiv:2007.03691] {10.1093/mnras/staa1977}, \href
  {https://ui.adsabs.harvard.edu/abs/2020MNRAS.tmp.2080W} {}

\bibitem[\protect\citeauthoryear{{Zabel} et~al.,}{{Zabel}
  et~al.}{2020}]{ZABE20}
{Zabel} N.,  et~al., 2020, \mn@doi [\mnras] {10.1093/mnras/staa1513}, \href
  {https://ui.adsabs.harvard.edu/abs/2020MNRAS.496.2155Z} {496, 2155}

\bibitem[\protect\citeauthoryear{da Silva, Fumagalli  \& Krumholz}{da~Silva
  et~al.}{2012}]{SILV12}
da Silva R.~L.,  Fumagalli M.,   Krumholz M.,  2012, \mn@doi [\apj]
  {10.1088/0004-637X/745/2/145}, \href
  {https://ui.adsabs.harvard.edu/abs/2012ApJ...745..145D} {745, 145}

\bibitem[\protect\citeauthoryear{da Silva, Fumagalli  \& Krumholz}{da~Silva
  et~al.}{2014}]{SILV14}
da Silva R.~L.,  Fumagalli M.,   Krumholz M.~R.,  2014, \mn@doi [\mnras]
  {10.1093/mnras/stu1688}, \href
  {https://ui.adsabs.harvard.edu/abs/2014MNRAS.444.3275D} {444, 3275}

\makeatother
\end{thebibliography}

	%%%%%%%%%%%%%%%%%%%%%%%%%%%%%%%%%%%%%%%%%%%%%%%%%%

	%%%%%%%%%%%%%%%%% APPENDICES %%%%%%%%%%%%%%%%%%%%%
	\appendix

	%%%%%%%%%%%%%%%%%%%%%%%%%%%%%%%%%%%%%%%%%%%%%%%%%%

	% Don't change these lines
	\bsp	% typesetting comment
	\label{lastpage}
\end{document}